\documentclass[amsmath,superscriptaddress,showpacs,prb,twocolumn] {revtex4-2}
\usepackage{bm}
\usepackage{bbm}
\usepackage{enumerate}
\usepackage{graphicx}
\usepackage[dvips]{epsfig}
\usepackage{epsf}
\usepackage{physics}
\usepackage{xcolor}
\usepackage{bbold}
\usepackage[normalem]{ulem}
\usepackage[caption=false]{subfig}
\usepackage{url}

\makeatletter
\newcommand{\Rmnum}[1]{\expandafter\@slowromancap\romannumeral #1@}
\makeatletter

\allowdisplaybreaks

\begin{document}
\title{Spin waves in layered antiferromagnets with honeycomb structure}
\author{Ankang Liu}
\affiliation{Department of Physics and Astronomy, Texas A\&M University, College Station, Texas 77843-4242, USA}
\author{Alexander M. Finkel'stein}
\affiliation{Department of Physics and Astronomy, Texas A\&M University, College Station, Texas 77843-4242, USA}
\affiliation{Department of Condensed Matter Physics, The Weizmann Institute of Science, Rehovot 76100, Israel}

\begin{abstract}

We develop a description of spin waves in a $3D$ quantum $XY$ antiferromagnet (AFM) in terms of macroscopic variables, magnetization and N\'eel vector densities. We consider a layered AFM with spins located on the honeycomb lattice. In the discussed system, the spectrum of spin waves consists of four modes, all well captured by our macroscopic description. The gapless mode of the spin waves, i.e., magnons, is described by a system of equations, which has a structure general for the Goldstone mode in AFMs. We demonstrate that the parameters in the spin Hamiltonian can be evaluated by fitting the experimental data with the results obtained for the four modes using the macroscopic variable approach. The description of AFM in terms of macroscopic variables can be easily extended to the case when the lattice of the magnetic substance is deformed by an external strain or acoustic wave.

\end{abstract}


\maketitle

\section{Introduction}\label{intro}
In this paper we derive the equations of motion for the system of spins in a quantum $XY$ AFM in terms of pairs of macroscopic quantities, which are the magnetization and the N\'eel vector densities. Compared to other classes, the $XY$ AFMs are relatively limited in occurrence. Usually, they could be met in systems with a hexagonal symmetry of the crystalline lattice, see Ref. [\onlinecite{oleaga20143d}]. At the present moment, a number of materials has been confirmed to be the layered AFM with in-plane spins, including NiPS$_3$ \cite{joy1992magnetism}, CoPS$_3$ \cite{wildes2017magnetic}, CuMnAs \cite{wadley2015antiferromagnetic}, CrCl$_3$ \cite{klein2019enhancement,cai2019atomically,kim2019evolution}, etc. Here, for concreteness, we consider the $XY$-type layered CoTiO$_3$, although the approach is general and expected to be applicable to any layered antiferromagnet. 

We arrive to a relatively simple description of quantum AFM in terms of the gradient expansion of the pairs of macroscopic variables that in the continuous limit reproduces the main features of the results obtained for CoTiO$_3$ in Ref. [\onlinecite{yuan2020dirac}]. In particular, this method allows to describe accurately all four spin-wave modes existing in the discussed system. By comparing their spectrum calculated here using  the macroscopic variables approach with the experimental data of Ref. [\onlinecite{yuan2020dirac}], we extracted the values of the parameters in the spin Hamiltonian and confirm the $XY$ character of the intralayer spin exchange in this material.

In a series of papers [\onlinecite{chumak2010all,karenowska2012oscillatory,chumak2017magnonic}], the magnon backward scattering by a magnonic crystal was studied experimentally in ferromagnets. The periodic scattering potential (i.e., the magnonic crystal) was created by a set of current carrying meander wires. The perspective of this experimental method for bulk AFM samples remains unclear. We, therefore, study here the effect of the lattice deformation on the spin dynamics. The deformations change distances between spins, and thus modify exchange coupling constants. The modulation of the coupling constants causes in its turn scattering of the spin waves. The description in terms of the macroscopic variables developed in this paper can be easily extended to a system with deformations, and allows one to obtain the dynamics of the scattering spin waves in the modulated crystal. This is another goal of the present paper.

\section{Spin Dynamics in the Absence of Lattice Deformations}\label{macro_and_free}
CoTiO$_3$ is a layered antiferromagnetic material, and is a sort of a magnetic ``ABC-stacked graphite".
\begin{figure}[htp] \centerline{\includegraphics[clip, width=1  \columnwidth]{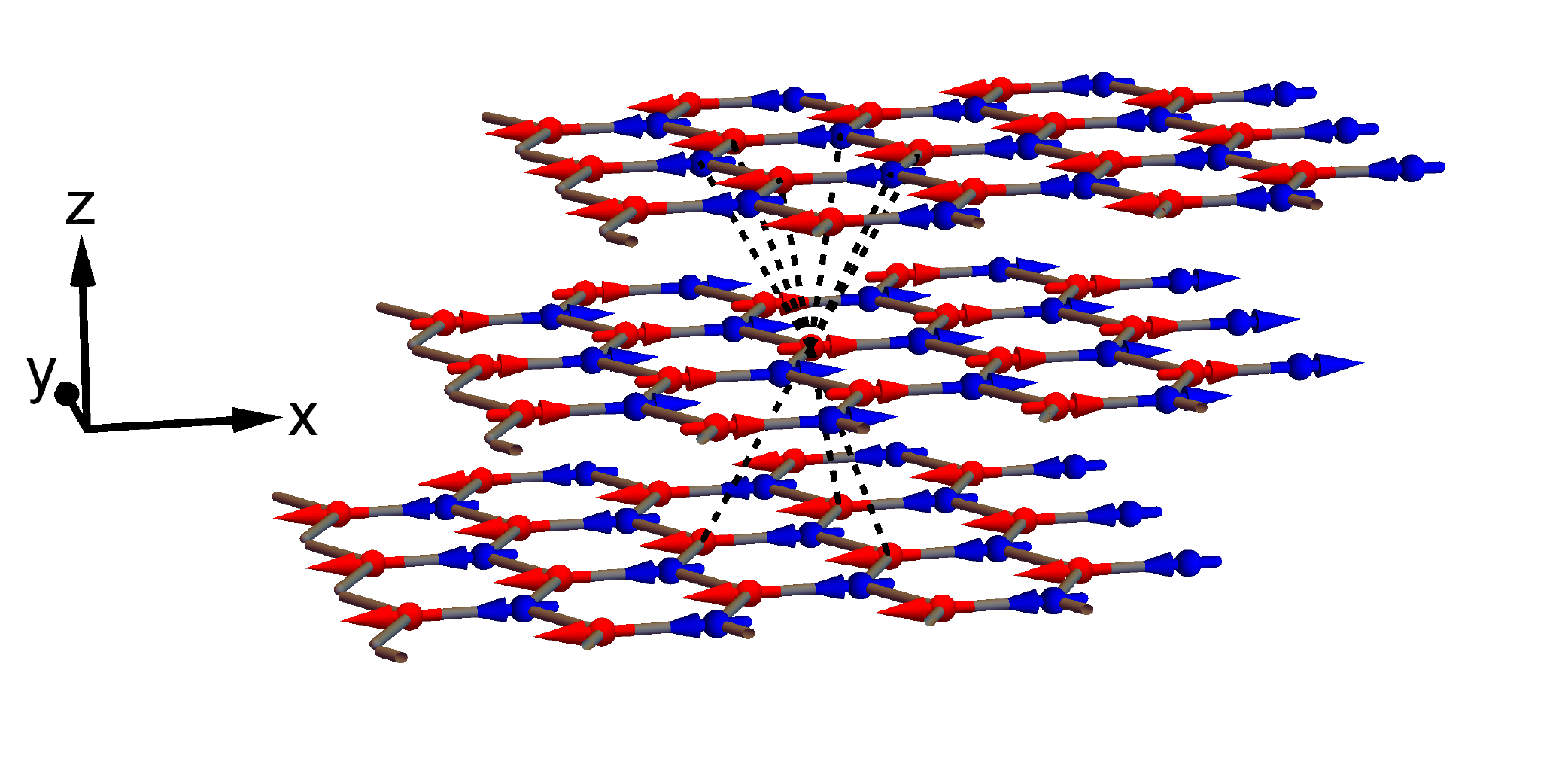}}
	
	\protect\caption{Schematic spin lattice structure of CoTiO$_3$. The period along the $z$ direction comprises six layers including the $ABC$ stacking and the alternating $\pm x$ spin ordering in the neighbor layers. The red and blue spheres represent atoms located on the $A$ and $B$ sublattices, respectively, while arrows indicate the direction of their spins. CoTiO$_3$ is an intralayer ferromagnet and simultaneously an interlayer antiferromagnet. The dashed lines display couplings of a selected atom in the middle layer to the nine next-nearest neighbors in the other two layers. Each of the red atoms on the $A$ sublattice in the middle layer is coupled with six blue atoms in the top layer and three red atoms in the bottom layer. A blue atom on the $B$ sublattice in the middle layer is coupled with three blue atoms in the top layer and six red atoms in the bottom layer.}
	
	\label{fig:spin_lattice}
	
\end{figure}
Namely, in each of the layers, spins are arranged on a honeycomb lattice ($xy$ plane), and different layers are ABC stacked along the third direction ($z$ axis). A schematic structure of the spin lattice is shown in Fig. \ref{fig:spin_lattice}. The exchange coupling $J_{\parallel}$ within a layer is ferromagnetic, i.e., $J_{\parallel}<0$, while the exchange coupling constant between layers is antiferromagnetic, $J_{\perp}>0$. Previous measurements (cf. Ref. [\onlinecite{yuan2020dirac}], and see also the discussion in Appendix \ref{XY_model}) found out that the Hamiltonian describing best the magnetic excitations in CoTiO$_3$ is
\begin{align}\label{H_spins_explicit}
	H=&\sum_{i,\delta_1}J_{\parallel}(S_{i}^{x}S_{i+\delta_1}^{x}+S_{i}^{y}S_{i+\delta_1}^{y})+\sum_{i,\delta_2}J_{\perp}(S_{i}^{x}\bar{S}_{i+\delta_2}^{x}
	\nonumber
	\\
	&+S_{i}^{y}\bar{S}_{i+\delta_2}^{y}+S_{i}^{z}\bar{S}_{i+\delta_2}^{z})+\{S^{x/y/z}\leftrightarrow\bar{S}^{x/y/z}\}.
\end{align}
Here, the index $i$ runs over all sites of spin, while $\delta_1$ and $\delta_2$ run over the nearest neighbors within the same layer, and all the next-nearest neighbors between the layers. In the Hamiltonian (\ref{H_spins_explicit}) we have introduced $\bm{S}$ and $\bar{\bm{S}}$ that are the spin operators for the $\pm x$-ordered magnetic layers, respectively. Using the Heisenberg equations of motion, $\frac{d\bm{S}}{dt}=\frac{1}{i}[\bm{S},H]$, for each of the spin components of $\bm{S}_i$, we find
\begin{align}\label{EOM_S}
	\frac{dS_{i}^{x}}{dt}=&\sum_{\delta_1}J_{\parallel}S_{i}^{z}S_{i+\delta_1}^{y}+\sum_{\delta_2}J_{\perp}(S_{i}^{z}\bar{S}_{i+\delta_2}^{y}-S_{i}^{y}\bar{S}_{i+\delta_2}^{z}),
	\nonumber
	\\
	\frac{dS_{i}^{y}}{dt}=&-\sum_{\delta_1}J_{\parallel}S_{i}^{z}S_{i+\delta_1}^{x}+\sum_{\delta_2}J_{\perp}(-S_{i}^{z}\bar{S}_{i+\delta_2}^{x}+S_{i}^{x}\bar{S}_{i+\delta_2}^{z}),
	\nonumber
	\\
	\frac{dS_{i}^{z}}{dt}=&\sum_{\delta_1}J_{\parallel}(S_{i}^{y}S_{i+\delta_1}^{x}-S_{i}^{x}S_{i+\delta_1}^{y})+\sum_{\delta_2}J_{\perp}(S_{i}^{y}\bar{S}_{i+\delta_2}^{x}
	\nonumber
	\\
	&-S_{i}^{x}\bar{S}_{i+\delta_2}^{y}).
\end{align}
The equations of motion for $\bar{\bm{S}}$ could be obtained by making the exchange: $\bm{S}\leftrightarrow\bar{\bm{S}}$. Although the period of the spin lattice along the $z$ direction is six layers, there is no need in considering the dynamics of all six layers. Instead, one just needs to calculate the equations of motion for $\bm{S}$ and $\bar{\bm{S}}$, which are the spin operators of two oppositely ordered layers. Suppose we start from the bottom layer in Fig. \ref{fig:spin_lattice} and move upward. Then each next layer requests for the same operation, i.e., change in the direction of the spin ordering along with a shift by one unit length along the $+x$ direction. Hence, each layer is in the same environment and does not feel the periodicity along the $z$ direction.

Next, in order to get a continuum model, we make an expansion in $\bm{S}$ with respect to its coordinate dependence, see e.g., Ref. [\onlinecite{mattis2012theory}]. For example, let us take the term $\sum_{\delta_1}J_{\parallel}S_{i}^{z}S_{i+\delta_1}^{y}$ and assume that the site $i$ is on the $A$ sublattice. We are interested in finding the equation of motion for a spin located on this site. For this, we need to explore its environment. Performing the expansion, we obtain
\begin{align}\label{gradient_expansion_1}
	\sum_{\delta_1}S_{i}^{z}S_{i+\delta_1}^{y}=&\sum_{\delta_1}S_{iA}^{z}(S_{iB}^{y}+\frac{\partial S_{iB}^{y}}{\partial r^{\alpha}}\delta_{1}^{\alpha}+\frac{1}{2}\frac{\partial^{2}S_{iB}^{y}}{\partial r^{\alpha}\partial r^{\beta}}\delta_{1}^{\alpha}\delta_{1}^{\beta}
	\nonumber
	\\
	&+\cdots)
	\nonumber
	\\
	\approx&3S_{iA}^{z}S_{iB}^{y}+\frac{3}{4}S_{iA}^{z}(\frac{\partial^{2}}{\partial x^{2}}+\frac{\partial^{2}}{\partial y^{2}})S_{iB}^{y}.
\end{align}
Here and further on, the subscription $A/B$ stands for the $A/B$ sublattices. There are three nearest neighbors for the honeycomb lattice, and $\bm{\delta}_{1}$ are taken to be $\bm{\delta}_{1,1}=(1,0,0)$, $\bm{\delta}_{1,2}=(-\frac{1}{2},\frac{\sqrt{3}}{2},0)$, and $\bm{\delta}_{1,3}=(-\frac{1}{2},-\frac{\sqrt{3}}{2},0)$. In Eq. (\ref{gradient_expansion_1}) the summation over $\alpha$ and $\beta$ is assumed, where $\alpha,\beta=x,y,z$ combines the three Cartesian components of the vector $\bm{\delta}_{1,j}$ with three coordinate derivatives. Note that for the convenience of the discussion, lengths are measured in the units of either intra- or interlayer lattice constants.

Similarly, for a term describing the interlayer interaction, $\sum_{\delta_2}J_{\perp}S_{i}^{z}\bar{S}_{i+\delta_2}^{y}$, we have
\begin{align}\label{gradient_expansion_2}
	\sum_{\delta_2}S_{i}^{z}\bar{S}_{i+\delta_2}^{y}\approx&3S_{iA}^{z}(\bar{S}_{iB}^{y}+2\bar{S}_{iA}^{y})+3S_{iA}^{z}[\frac{\partial\bar{S}_{iB}^{y}}{\partial z}
	\nonumber
	\\
	&+\frac{1}{4}(\frac{\partial^{2}}{\partial x^{2}}+\frac{\partial^{2}}{\partial y^{2}}+2\frac{\partial^{2}}{\partial z^{2}})(\bar{S}_{iB}^{y}+2\bar{S}_{iA}^{y})].
\end{align}
Here we assumed that a site $i$ is on one of the $A$ sublattices (i.e., $i$ is red), and that $\delta_{2}$ runs over nine next-nearest neighbors as indicated by the dashed line in Fig. \ref{fig:spin_lattice}.

To derive the equations of motion for the macroscopic quantities, we substitute the leading expansion terms as shown in Eqs. (\ref{gradient_expansion_1}) and (\ref{gradient_expansion_2}) back into Eq. (\ref{EOM_S}). The calculations are straightforward, and details are presented in Appendix \ref{derivation_without_deformation}. Then, after making a transition from the site spin operators $\bm{S}_{i}(t)$ to the continuous variable $\bm{S}(\bm{r},t)$, we introduce the macroscopic quantities for each of the two sublattices: the total magnetization $\bm{m}_{A/B}(\bm{r},t)\equiv\bm{S}_{A/B}(\bm{r},t)+\bar{\bm{S}}_{A/B}(\bm{r},t)$ and the N\'eel vector $\bm{l}_{A/B}(\bm{r},t)\equiv\bm{S}_{A/B}(\bm{r},t)-\bar{\bm{S}}_{A/B}(\bm{r},t)$, which will be used for describing the long-wavelength spin wave excitation. Finally, for the two spin-wave branches with the lowest energy we implement the approximation $\bm{m}_{A}=\bm{m}_{B}=\bm{m}$ and $\bm{l}_{A}=\bm{l}_{B}=\bm{l}$, and as a result get:
\begin{align}\label{EOM_m_1}
	\frac{dm^{x}}{dt}\approx\frac{3}{2}J_{\parallel}(l^{z}l^{y})+\frac{9}{8}J_{\perp}(l^{y}\nabla_{+}^{2}l^{z}),
\end{align}
\begin{align}\label{EOM_m_2}
	\frac{dm^{y}}{dt}\approx-\frac{3}{2}J_{\parallel}(l^{z}l^{x})-\frac{9}{8}J_{\perp}(l^{x}\nabla_{+}^{2}l^{z}),
\end{align}
\begin{align}\label{EOM_m_3}
	\frac{dm^{z}}{dt}\approx-\frac{3}{8}J_{\parallel}[\bm{l}\times(\nabla_{-}^{2}\bm{l})]_{z}+\frac{9}{8}J_{\perp}[\bm{l}\times(\nabla_{+}^{2}\bm{l})]_{z},
\end{align}
\begin{align}\label{EOM_l_1}
	\frac{dl^{x}}{dt}\approx(\frac{3}{2}J_{\parallel}-9J_{\perp})(l^{y}m^{z})-\frac{9}{8}J_{\perp}(l^{y}\nabla_{+}^{2}m^{z}),
\end{align}
\begin{align}\label{EOM_l_2}
	\frac{dl^{y}}{dt}\approx(-\frac{3}{2}J_{\parallel}+9J_{\perp})(l^{x}m^{z})+\frac{9}{8}J_{\perp}(l^{x}\nabla_{+}^{2}m^{z}),
\end{align}
and
\begin{align}\label{EOM_l_3}
	\frac{dl^{z}}{dt}\approx&9J_{\perp}(\bm{m}\times\bm{l})_{z}-\frac{3}{8}J_{\parallel}[\bm{l}\times(\nabla_{-}^{2}\bm{m})]_{z}
	\nonumber
	\\
	&-\frac{9}{8}J_{\perp}[\bm{l}\times(\nabla_{+}^{2}\bm{m})]_{z}.
\end{align}
Here, the short notation $\nabla_{\pm}^{2}\equiv\nabla^{2}\pm\frac{\partial^{2}}{\partial z^{2}}$ has been introduced; the indices  $x,y,z$ mark the $x$, $y$, and $z$ component of the vectors, respectively. In the following, we drop the last terms in Eqs. (\ref{EOM_l_1}) and (\ref{EOM_l_2}), because they lead to the terms in dispersion, which are of higher order in $k^2$, while we are interested in $\omega^2$ only up to the order $\sim\mathcal{O}(k^2)$.

To proceed further, we will use the parametrization (see e.g., Refs. [\onlinecite{haldane1983nonlinear,auerbach2012interacting,takei2014superfluid}])
\begin{align}\label{parametrization}
	\bm{l}=&2\tilde{S}\sqrt{1-\left(\frac{|\bm{m}|}{2\tilde{S}}\right)^2}\left(\cos\theta\cos\phi,\cos\theta\sin\phi,\sin\theta\right),
	\nonumber
	\\
	\bm{m}=&(\allowbreak-m_{\theta}\sin\theta\cos\phi-m_{\phi}\sin\phi,-m_{\theta}\sin\theta\sin\phi+m_{\phi}
	\nonumber
	\\
	&\times\cos\phi,m_{\theta}\cos\theta).
\end{align}
Under this parametrization, the vectors $\bm{m}$ and $\bm{l}$ are automatically constrained by $\bm{m}\cdot\bm{l}=0$ and, at the same time, the lengths of $\bm{S}$ and $\bar{\bm{S}}$ are taken to be a fixed value $\tilde{S}$. De facto, by the transition from spin operators to the classical variables $\bm{m}$ and $\bm{l}$, we have implemented the language of the nonlinear sigma model (NLSM) for the description of the antiferromagnet dynamics.

Keeping only the linear terms in Eqs. (\ref{EOM_m_1})--(\ref{EOM_l_3}), we get two decoupled pairs of equations in terms of the variables introduced in Eqs. (\ref{parametrization}):
\begin{align}\label{EOM_m_theta}
	&\dot{m}_{\theta}\approx(4\tilde{S}^{2})(-\frac{3}{8}J_{\parallel}\nabla_{-}^{2}\phi+\frac{9}{8}J_{\perp}\nabla_{+}^{2}\phi),
	\nonumber
	\\
	&\dot{\phi}\approx(-\frac{3}{2}J_{\parallel}+9J_{\perp})m_{\theta};
\end{align}
and
\begin{align}\label{EOM_m_phi}
	&\dot{m}_{\phi}\approx(4\tilde{S}^{2})(-\frac{3}{2}J_{\parallel}\theta-\frac{9}{8}J_{\perp}\nabla_{+}^{2}\theta),
	\nonumber
	\\
	&\dot{\theta}\approx(-9J_{\perp}-\frac{3}{8}J_{\parallel}\nabla_{-}^{2}-\frac{9}{8}J_{\perp}\nabla_{+}^{2})m_{\phi}.
\end{align}
Note that $\theta^{(0)}=0$, $m_{\theta}^{(0)}=0$, and $m_{\phi}^{(0)}=0$ are the equilibrium values for these equations, while $\phi$ can be arbitrary, because this system has a rotational symmetry along the $z$ direction.

By taking another time derivative in Eqs. (\ref{EOM_m_theta}) and (\ref{EOM_m_phi}) we obtain closed equations of the second order. For example, for ${m_{\theta}}$ and ${m_{\phi}}$ they look as follows:
\begin{align}\label{eigen_eqs}
	\ddot{m}_{\theta}\approx&(4\tilde{S}^{2})[-\frac{3J_{\parallel}}{8}(-\frac{3J_{\parallel}}{2}+9J_{\perp})\nabla_{-}^{2}+\frac{9J_{\perp}}{8}(-\frac{3J_{\parallel}}{2}
	\nonumber
	\\
	&+9J_{\perp})\nabla_{+}^{2}]m_{\theta},
	\nonumber
	\\
	\ddot{m}_{\phi}\approx&(4\tilde{S}^{2})[\frac{27}{2}J_{\parallel}J_{\perp}+\frac{9}{16}J_{\parallel}^{2}\nabla_{-}^{2}+\frac{9J_{\perp}}{8}(\frac{3J_{\parallel}}{2}
	\nonumber
	\\
	&+9J_{\perp})\nabla_{+}^{2}]m_{\phi}.
\end{align}
These equations give eigenfrequencies of the two low-energy spin-wave branches
\begin{align}\label{eigen_values}
	\omega_a\approx&(2\tilde{S})[(\frac{9}{16}J_{\parallel}^{2}-\frac{81}{16}J_{\parallel}J_{\perp}+\frac{81}{8}J_{\perp}^{2})(k_{x}^{2}+k_{y}^{2})
	\nonumber
	\\
	&+(-\frac{27}{8}J_{\parallel}J_{\perp}+\frac{81}{4}J_{\perp}^{2})k_{z}^{2}]^{\frac{1}{2}},
	\nonumber
	\\
	\omega_{sh}\approx&(2\tilde{S})[-\frac{27}{2}J_{\parallel}J_{\perp}+\frac{9}{16}J_{\parallel}^{2}(k_{x}^{2}+k_{y}^{2})	\nonumber
	\\
	&+(\frac{81}{8}J_{\perp}^{2}+\frac{27}{16}J_{\parallel}J_{\perp})(k_{x}^{2}+k_{y}^{2}+2k_{z}^{2})]^{\frac{1}{2}}.
\end{align}
Here, $\omega_a$ and $\omega_{sh}$ are the acousticlike and the opticlike branches of the spin waves, respectively. In fact, the two branches exactly repeat each other after \emph{shifting} $k_z$ on $\pm\pi$. (This is why we indicate the ``fake" opticlike branch by ``$sh$".)

It remains to obtain the two ``true" opticlike branches with higher energies. Since the opticlike excitations are not related with the rotational symmetry along the $z$ directions, we perturb the spins on $A$ and $B$ sublattices in the anti-phase manner: An ansatz $\bm{m}_{A/B}=\pm(\delta m^x\bm{e}_x+\delta m^y\bm{e}_y+\delta m^z\bm{e}_z)$ and $\bm{l}_{A/B}=2\tilde{S}\bm{e}_x\pm(\delta l^x\bm{e}_x+\delta l^y\bm{e}_y+\delta l^z\bm{e}_z)$ is implemented for these eigenmodes. Here, $+$ and $-$ stand for $A$ and $B$ sublattices, respectively. Note that, without loss of generality, we take the equilibrium N\'eel vector to be along the $x$ direction, $\bm{l}_0=2\tilde{S}\bm{e}_x$. As above, $\bm{m}_{A/B}$ and $\bm{l}_{A/B}$ are subject to the constraint $\bm{m}_{A/B}\cdot\bm{l}_{A/B}=0$ \cite{note1}. Expanding the magnetization density and N\'eel vector around the equilibrium, we get (more details can be found in Appendix \ref{derivation_without_deformation})
\begin{align}\label{EOM_m_x} 
	\frac{d\delta m^{x}}{dt}\approx0,
\end{align}
\begin{align}\label{EOM_m_y}
	\frac{d\delta m^{y}}{dt}\approx2\tilde{S}[(-\frac{3}{2}J_{\parallel}+3J_{\perp})-\frac{3}{8}J_{\perp}\nabla_{+}^{2}]\delta l^{z},
\end{align}
\begin{align}\label{EOM_m_z}
	\frac{d\delta m^{z}}{dt}\approx2\tilde{S}[(3J_{\parallel}-3J_{\perp})+\frac{3}{8}J_{\parallel}\nabla_{-}^{2}+\frac{3}{8}J_{\perp}\nabla_{+}^{2}]\delta l^{y},
\end{align}
\begin{align}\label{EOM_l_x}
	\frac{d\delta l^{x}}{dt}\approx0,
\end{align}
\begin{align}\label{EOM_l_y}
	\frac{d\delta l^{y}}{dt}\approx2\tilde{S}[(-\frac{3}{2}J_{\parallel}+6J_{\perp})+\frac{3}{8}J_{\perp}\nabla_{+}^{2}]\delta m^{z},
\end{align}
and
\begin{align}\label{EOM_l_z}
	\frac{d\delta l^{z}}{dt}\approx2\tilde{S}[(3J_{\parallel}-6J_{\perp})+\frac{3}{8}J_{\parallel}\nabla_{-}^{2}-\frac{3}{8}J_{\perp}\nabla_{+}^{2}]\delta m^{y}.
\end{align}
We, thus, get two pairs of equations: Eqs. (\ref{EOM_m_y}) and (\ref{EOM_l_z}) for the pair $(\delta m^y,\delta l^z)$, and Eqs. (\ref{EOM_m_z}) and (\ref{EOM_l_y}) for $(\delta m^z,\delta l^y)$. Consequently, these pairs of equations lead us to two opticlike modes:
\begin{align}\label{eigen_values_higher}
	\omega_{o1}\approx&(2\tilde{S})[\frac{9}{2}(J_{\parallel}-2J_{\perp})^2-\frac{9}{16}J_{\parallel}(J_{\parallel}-2J_{\perp})(k_{x}^{2}+k_{y}^{2})
	\nonumber
	\\
	&-\frac{9}{16}J_{\perp}(J_{\parallel}-2J_{\perp})(k_{x}^{2}+k_{y}^{2}+2k_{z}^{2})]^{\frac{1}{2}},
	\nonumber
	\\
	\omega_{o2}\approx&(2\tilde{S})[\frac{9}{2}(J_{\parallel}-J_{\perp})(J_{\parallel}-4J_{\perp})-\frac{9}{16}J_{\parallel}(J_{\parallel}-4J_{\perp})
	\nonumber
	\\
	&\times(k_{x}^{2}+k_{y}^{2})+\frac{9}{16}J_{\perp}(J_{\parallel}+2J_{\perp})(k_{x}^{2}+k_{y}^{2}+2k_{z}^{2})]^{\frac{1}{2}}.
\end{align}
The gradient terms in Eqs. (\ref{EOM_m_y}),  (\ref{EOM_l_z}), and in (\ref{EOM_m_z}), (\ref{EOM_l_y}) determine the dispersion of the opticlike modes.

\section{Holstein–Primakoff Approach}\label{H_P}
As a comparison, we introduce an 8$\times$8 model using the Holstein–Primakoff transformation, which quantitatively describes the spectrum of the spin waves in CoTiO$_3$ with using the Hamiltonian given by Eq. (\ref{H_spins_explicit}).

\subsection{The 8$\times$8 model}\label{8_times_8}
For a layer, where the magnetization is ordered along $x$ direction, we introduce the standard Holstein-Primakoff operators
\begin{align}\label{H-P_transformations_1}
	&S^{x}_{A/B}=\tilde{S}-(a^{\dagger}a)/(b^{\dagger}b),
	\nonumber
	\\
	&S^{+}_{A/B}\equiv S^{y}_{A/B}+iS^{z}_{A/B}=\sqrt{2\tilde{S}-(a^{\dagger}a/b^{\dagger}b)}(a/b),
	\nonumber
	\\
	&S^{-}_{A/B}\equiv S^{y}_{A/B}-iS^{z}_{A/B}=(a^{\dagger}/b^{\dagger})\sqrt{2\tilde{S}-(a^{\dagger}a/b^{\dagger}b)}.
\end{align}
Here, the subscription $A/B$ indicates the $A/B$ sublattices and, similarly, $a^{\dagger}/b^{\dagger}$ and $a/b$ are creation and annihilation operators of spin excitations on the $A$ and $B$ sublattices, respectively. In the discussed system, CoTiO$_3$, the effective spin $\tilde{S}=1/2$; see the discussion on this point in Ref. [\onlinecite{yuan2020dirac}]. Finally, for the neighboring layer, where the magnetization is ordered along the $-x$ direction, we use operators marked with a bar. We have
\begin{align}\label{H-P_transformations_2}
	&\bar{S}^{x}_{A/B}=(\bar{a}^{\dagger}\bar{a}/\bar{b}^{\dagger}\bar{b})-\tilde{S},
	\nonumber
	\\
	&\bar{S}^{+}_{A/B}\equiv -\bar{S}^{y}_{A/B}+i\bar{S}^{z}_{A/B}=\sqrt{2\tilde{S}-(\bar{a}^{\dagger}\bar{a}/\bar{b}^{\dagger}\bar{b})}(\bar{a}/\bar{b}),
	\nonumber
	\\
	&\bar{S}^{-}_{A/B}\equiv -\bar{S}^{y}_{A/B}-i\bar{S}^{z}_{A/B}=(\bar{a}^{\dagger}/\bar{b}^{\dagger})\sqrt{2\tilde{S}-(\bar{a}^{\dagger}\bar{a}/\bar{b}^{\dagger}\bar{b})}.
\end{align}

Keeping only the quadratic form in terms of the creation and annihilation operators, one obtains a Hamiltonian $H_{SW}$ in the quasimomentum $\bm{k}$ space, which determines the spectrum of the spin waves. The Hamiltonian $H_{SW}=\tilde{S}\sum_{k}V_{k}^{\dagger}H_{k}V_{k}$ is determined as follows:
\begin{align}\label{V_k}
	V_{k}=\{a_k,b_k,a_{-k}^{\dagger},b_{-k}^{\dagger},\bar{a}_k,\bar{b}_k,\bar{a}_{-k}^{\dagger},\bar{b}_{-k}^{\dagger}\}^T
\end{align}
and
\begin{align}\label{H_k}
	H_k=
	\begin{pmatrix}
		H_{1k}&H_{2k}\\
		H_{2k}&H_{1k}\\
	\end{pmatrix}.
\end{align}
Here, $H_{1k}$ and $H_{2k}$ are $4\times4$ matrices
\begin{align}\label{H_1}
	H_{1k}=
	\begin{pmatrix}
		A_{k}&B_{k}&0&B_{k}\\
		B_{k}^{*}&A_{k}&B_{k}^{*}&0\\
		0&B_{k}&A_{k}&B_{k}\\
		B_{k}^{*}&0&B_{k}^{*}&A_{k}\\
	\end{pmatrix}
\end{align}
and
\begin{align}\label{H_2}
	H_{2k}=
	\begin{pmatrix}
		0&0&C_{k}&F_{k}\\
		0&0&F_{k}^{*}&C_{k}\\
		C_{k}&F_{k}&0&0\\
		F_{k}^{*}&C_{k}&0&0\\
	\end{pmatrix}.
\end{align}
The matrix elements here are $A_k=-3J_{\parallel}+9J_{\perp}$, $B_k=\frac{1}{2}J_{\parallel}\gamma_k$, $C_k=-J_{\perp}(e^{-ik_z}\gamma_k+e^{ik_z}\gamma_{k}^{*})$, and $F_k=-J_{\perp}e^{-ik_z}\gamma_{k}^{*}$. The factor $\gamma_k$ is determined by summation over the nearest neighbors, i.e., for the honeycomb lattice $\gamma_k=\sum_{j=1,2,3}e^{i\bm{k}\cdot\bm{\delta}_{1,j}}$ with $\bm{\delta}_{1,1}=(1,0,0)$, $\bm{\delta}_{1,2}=(-\frac{1}{2},\frac{\sqrt{3}}{2},0)$, and $\bm{\delta}_{1,3}=(-\frac{1}{2},-\frac{\sqrt{3}}{2},0)$. In our discussions, we take both in-plane and out-of-plane lattice constants to be $1$ for simplicity. In Ref. [\onlinecite{yuan2020dirac}], the best estimates of $J_{\parallel}$ and $J_{\perp}$, which match quantitatively well with the experimental data are found to be $J_{\parallel}=-4.41$ meV and $J_{\perp}=0.57$ meV. In the discussions below, we will use for the parameters $J_{\parallel}$ and $J_{\perp}$ these values.
	
In Appendix \ref{XY_model}, we get the parameters of the Hamiltonian (\ref{H_spins_explicit}) by analyzing the experimental data from Ref. [\onlinecite{yuan2020dirac}] using our macroscopic description developed above. The extracted values of parameters, which optimally fit the data, are very close to the ones presented in Ref. [\onlinecite{yuan2020dirac}].

Note that, this $8\times8$ model gives $4$ branches of the magnon spectrum. These $4$ branches could be divided into $2$ groups by the symmetry of the eigenstates. To find the spectrum of magnons, one needs to solve the eigenvalue problem $H_k\ket{\psi}=E_k S_3\ket{\psi}$ with the diagonal matrix $S_3=$ diag$(1,1,-1,-1,1,1,-1,-1)$. The eigen vector $\ket{\psi}$ here is an eight-dimensional vector constructed in the basis $V_k$, see Eq. (\ref{V_k}). It could be written as $\ket{\psi}=\{\psi_{1}^{T},\psi_{2}^{T}\}^T$, where $\psi_1$ and $\psi_2$ are four-dimensional vectors within the subspaces $\{a_k,b_k,a_{-k}^{\dagger},b_{-k}^{\dagger}\}^T$ and $\{\bar{a}_k,\bar{b}_k,\bar{a}_{-k}^{\dagger},\bar{b}_{-k}^{\dagger}\}^T$, respectively. For one group of the eigenstates, which has the property $\psi_1=\psi_2$, the eigenvalue equation becomes $(H_{1k}+H_{2k})\psi_1=E_k\sigma_3\psi_1$, where the diagonal $\sigma_3=$ diag$(1,1,-1,-1)$ is a $4\times4$ matrix. For another group of the eigenstates with the property $\psi_1=-\psi_2$, the eigenvalue equation reduces to $(H_{1k}-H_{2k})\psi_1=E_k\sigma_3\psi_1$. The effective $4\times4$ Hamiltonian $H_{1k}+H_{2k}$ coincides with Eqs. (6) and (7) in the Supplemental Material of Ref. [\onlinecite{yuan2020dirac}]. Each of the reduced $4\times4$ Hamiltonians, $H_{1k}\pm H_{2k}$, describes two branches of the spin waves. 

\begin{figure}[htp] \centerline{\includegraphics[clip, width=1  \columnwidth]{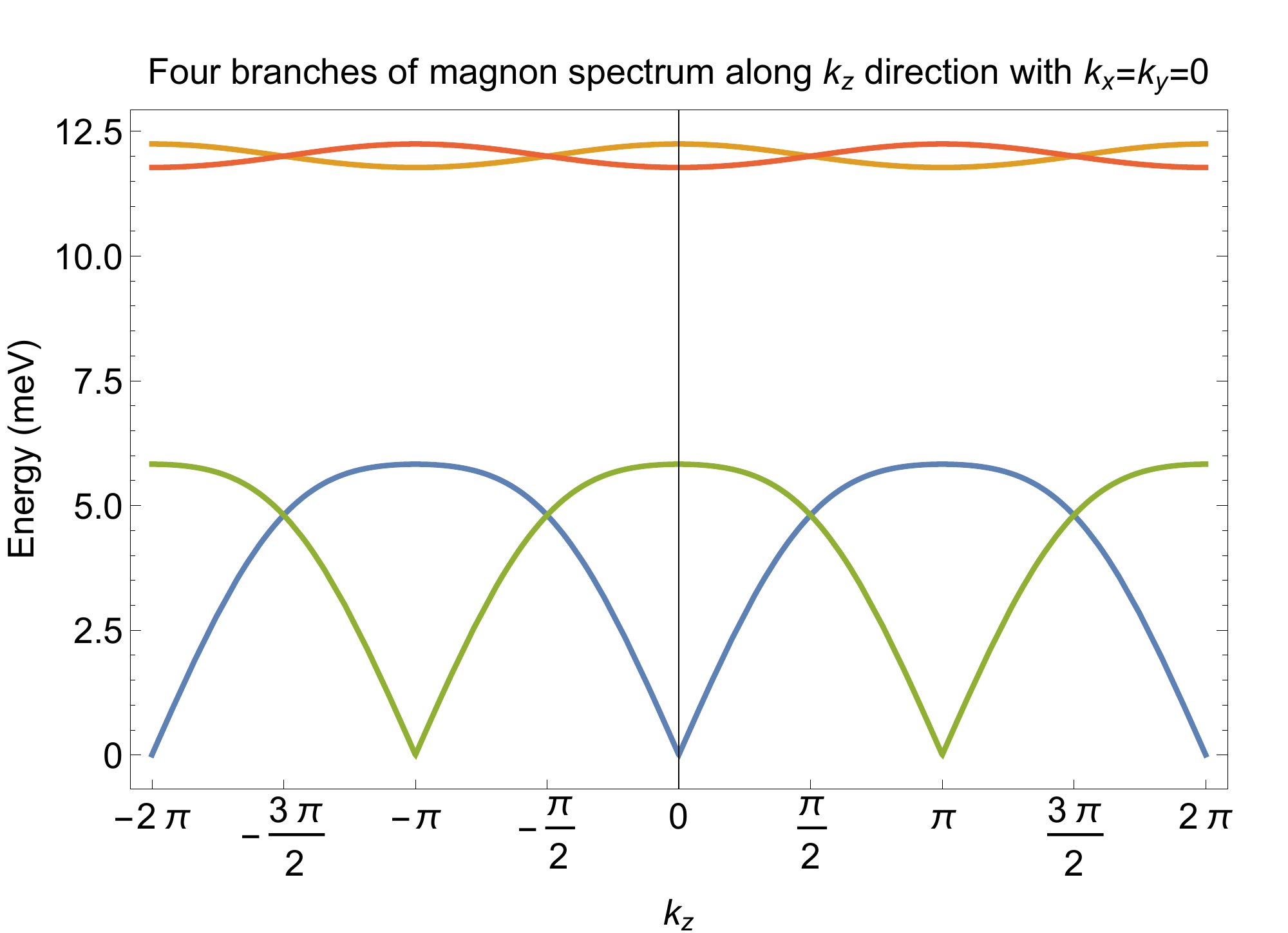}}		
	\protect\caption{Four branches of the magnon spectrum obtained by solving the equations $(H_{1k}\pm H_{2k})\psi_1=E_k\sigma_3\psi_1$ with $J_{\parallel}=-4.41$ meV and $J_{\perp}=0.57$ meV. The eigenvalues are plotted along the $k_z$ direction with $k_x=k_y=0$.}		
	\label{fig:four_branches}		
\end{figure}

We plot in Fig. \ref{fig:four_branches} the whole magnon spectrum consisting of four branches. The blue and orange curves are derived from $(H_{1k}+H_{2k})\psi_1=E_k\sigma_3\psi_1$, while green and red ones correspond to $(H_{1k}-H_{2k})\psi_1=E_k\sigma_3\psi_1$. Actually, these two pairs of branches are connected through $\pi$ shifting along the $k_z$ direction. This reveals the additional symmetry $\psi_1=\pm\psi_2$ possessed by a spin system with the layered structure of CoTiO$_3$.  

One of the four branches, the blue curve, touches zero at $\bm{k}=0$. This acousticlike branch corresponds to the Goldstone mode. It is a direct consequence of the continuous symmetry with respect to rotation of the N\'eel vector $\bm{l}$ in the $xy$ plane. The other branch (the green curve) after shifting $k_z$ by $\pi$ reproduces the Goldstone mode.

\subsection{The eigenstates and eigenfrequencies}\label{H_P_approxiamtion}
In principle, the magnon spectrum as well as its eigenstates can be found by solving the eigenvalue equations $(H_{1k}\pm H_{2k})\psi_1=E_{k}^{(\pm)}\sigma_3\psi_1$. However, it is very intractable, and therefore we present an approximation here, which allows us to describe the eigenstates and eigenfrequencies in a simplified but still comprehensive way. As an example, we demonstrate how it works for $H^{+}\equiv H_{1k}+H_{2k}$. Under the basis of the subspace $\{a_k,a_{-k}^{\dagger},b_k,b_{-k}^{\dagger}\}^T$, $H^{+}$ becomes
\begin{align}\label{H_plus} 
	H^{+}=
	\begin{pmatrix}
		A_{k}&C_{k}&B_{k}&G_{k}^{+}\\
		C_{k}&A_{k}&G_{k}^{+}&B_{k}\\
		B_{k}^{*}&(G_{k}^{+})^{*}&A_{k}&C_{k}\\
		(G_{k}^{+})^{*}&B_{k}^{*}&C_{k}&A_{k}\\
	\end{pmatrix}
\end{align}
with $G_{k}^{+}\equiv B_{k}+F_{k}$. Since the $A$ and $B$ sublattices are equivalent, we are looking for the eigenstate in the form
\begin{align}\label{ansatz_state}
	\tilde{\psi}_{1}=a
	\begin{pmatrix}
		e^{i\chi_{a}}\\
		1\\
		e^{i\chi_{b_1}}\\
		e^{i\chi_{b_2}}
	\end{pmatrix}
    +\frac{1}{8a}
    \begin{pmatrix}
    	e^{i\chi_{a}}\\
    	-1\\
    	e^{i\chi_{b_1}}\\
    	-e^{i\chi_{b_2}}
    \end{pmatrix}.
\end{align}
Here, $a$, $\chi_{a}$, $\chi_{b_1}$, and $\chi_{b_2}$ are functions of the wave vector $\bm{k}$ that need to be evaluated. The magnitudes of the $\tilde{\psi}$ components are all determined by the parameter $a$, and are equal to $a\pm{\frac{1}{8a}}$. Note that $\tilde{\psi}_{1}$ satisfies the standard normalization condition $\tilde{\psi}_{1}^{\dagger}\tilde{\sigma}_{3}\tilde{\psi}_{1}=1$, where $\tilde{\sigma}_{3}=$ diag$(1,-1,1,-1)$. Equation (\ref{ansatz_state}) states that the dynamics of spins on the $B$ sublattice is the same as the one on the $A$ sublattice, except for the phase difference.

Expanding $\chi_{a}$, $\chi_{b_1}$, and $\chi_{b_2}$ around $0$ (see Appendix \ref{eqs_for_ansatz}), we get the approximate solution of the equation $H^{+}\tilde{\psi}_1=E\tilde{\sigma}_{3}\tilde{\psi}_1$:
\begin{widetext}
\begin{align}\label{solution_to_eigenstates_1}
	E\approx&\sqrt{[A_{k}+\Re(B_{k})+C_{k}+\Re(G_{k}^{+})][A_{k}+\Re(B_{k})-C_{k}-\Re(G_{k}^{+})]},
	\nonumber
	\\
	a\approx&\frac{1}{2\sqrt{2}}\Big(\frac{A_{k}+\Re(B_{k})-C_{k}-\Re(G_{k}^{+})}{A_{k}+\Re(B_{k})+C_{k}+\Re(G_{k}^{+})}\Big)^{\frac{1}{4}},
\end{align}
and
\begin{align}\label{solution_to_eigenstates_2}
	\chi_{a}\approx&-\frac{16a^{2}[\Im(G_{k}^{+})\Re(B_{k})-\Im(B_{k})\Re(G_{k}^{+})]}{\Re(B_{k})[(64a^{4}-1)\Re(B_{k})+(64a^{4}+1)\Re(G_{k}^{+})]+C_{k}[(64a^{4}+1)\Re(B_{k})+(64a^{4}-1)\Re(G_{k}^{+})]},
	\nonumber
	\\
	\chi_{b_1}\approx&-\frac{\Re(B_{k})[(64a^{4}-1)\Im(B_{k})+(64a^{4}+1)\Im(G_{k}^{+})]+C_{k}[(64a^{4}+1)\Im(B_{k})+(64a^{4}-1)\Im(G_{k}^{+})]}{\Re(B_{k})[(64a^{4}-1)\Re(B_{k})+(64a^{4}+1)\Re(G_{k}^{+})]+C_{k}[(64a^{4}+1)\Re(B_{k})+(64a^{4}-1)\Re(G_{k}^{+})]},
	\nonumber
	\\
	\chi_{b_2}=&\chi_{a}+\chi_{b_1}
\end{align}
\end{widetext}
where $\Re(\cdots)$ and $\Im(\cdots)$ denote the real and imaginary part of ``$\cdots$'', respectively. The solution presented by Eq. (\ref{solution_to_eigenstates_2}) indicates the smallness of the phases, which is consistent with the expansion in phases $\chi_{a}$, $\chi_{b_1}$, and $\chi_{b_2}$ performed after Eq. (\ref{ansatz_state}).

\begin{figure*}[htp]
	\subfloat[Spectrum of the acousticlike branch \label{subfig:exact_vs_approximated_1}]{
		\includegraphics[width=0.33 \textwidth]{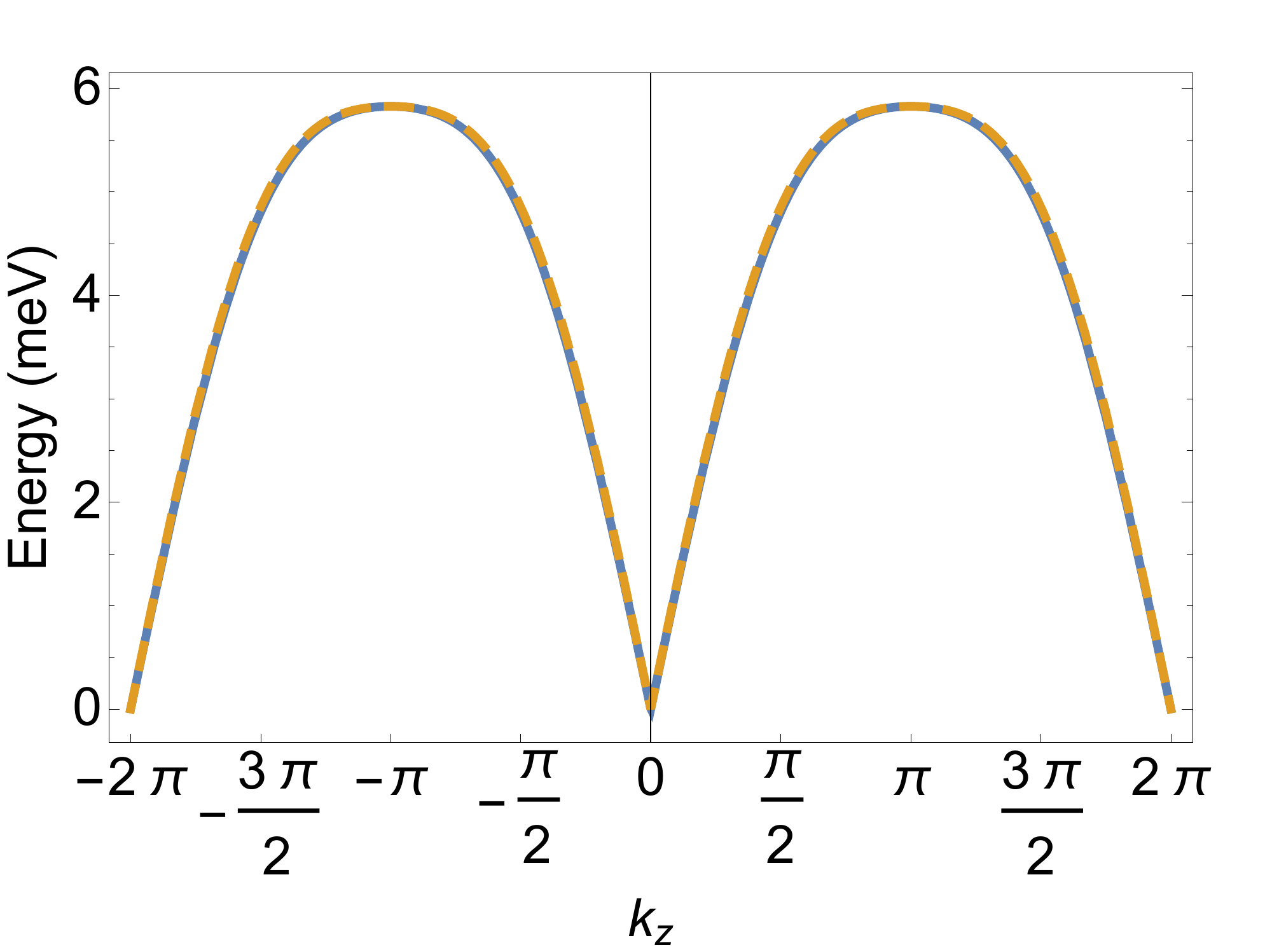}
	}
	\subfloat[Component phases of the acousticlike magnon state \label{subfig:exact_vs_approximated_2}]{
		\includegraphics[width=0.33 \textwidth]{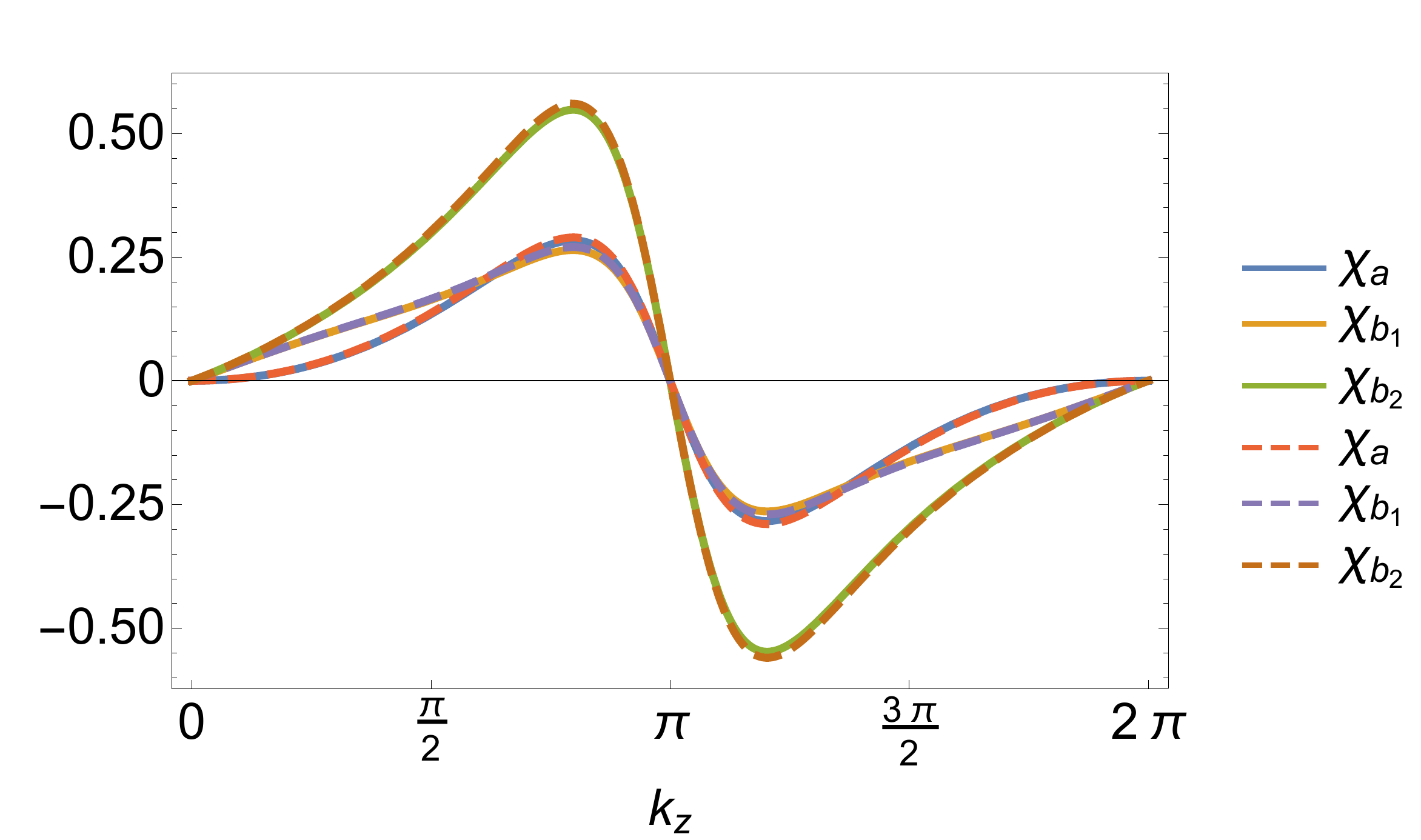}
	}
	\subfloat[Amplitudes on each component of the acousticlike magnon state \label{subfig:exact_vs_approximated_3}]{
		\includegraphics[width=0.33 \textwidth]{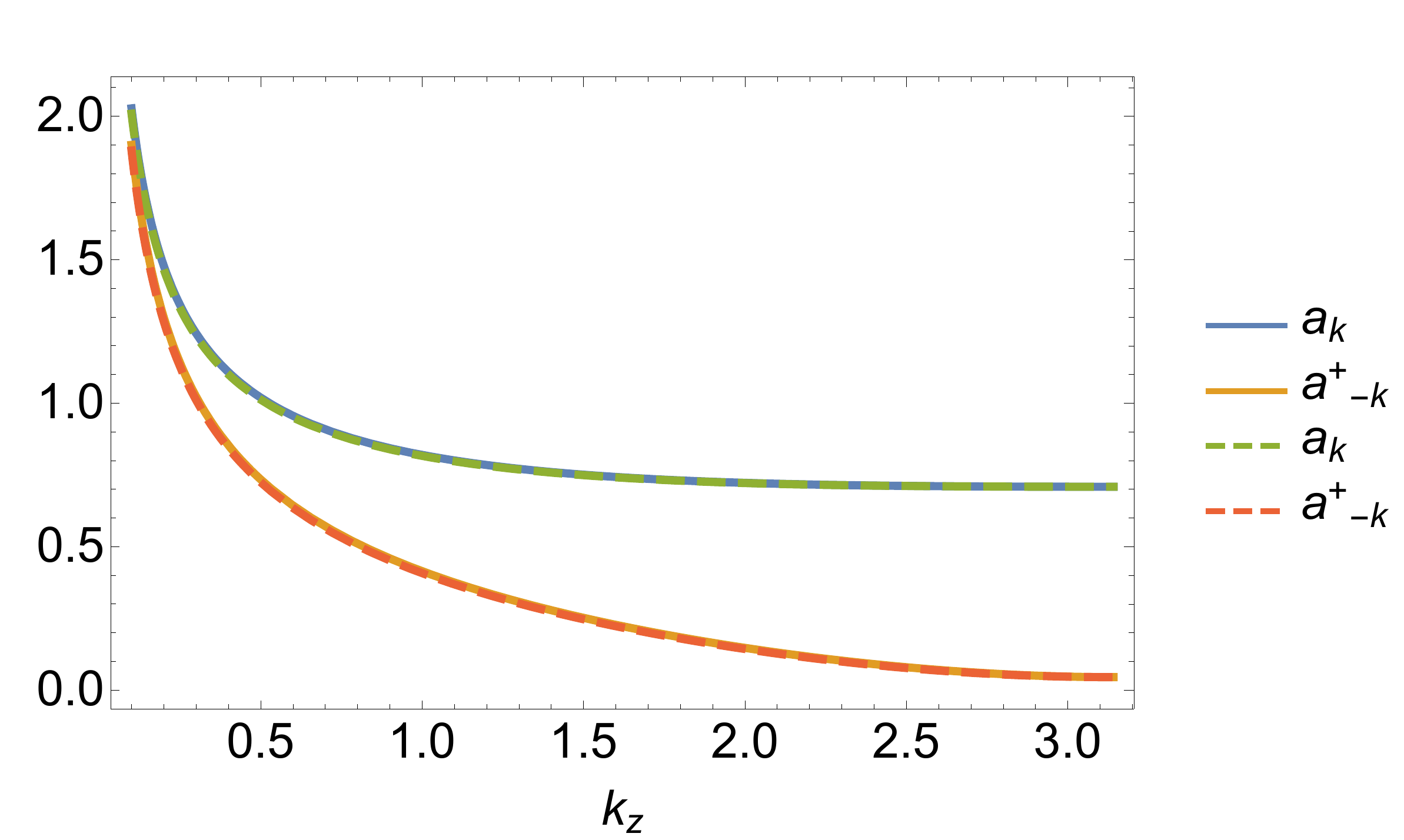}
	}\\
	
	\subfloat[Spectrum of the opticlike branch \label{subfig:exact_vs_approximated_4}]{
		\includegraphics[width=0.33 \textwidth]{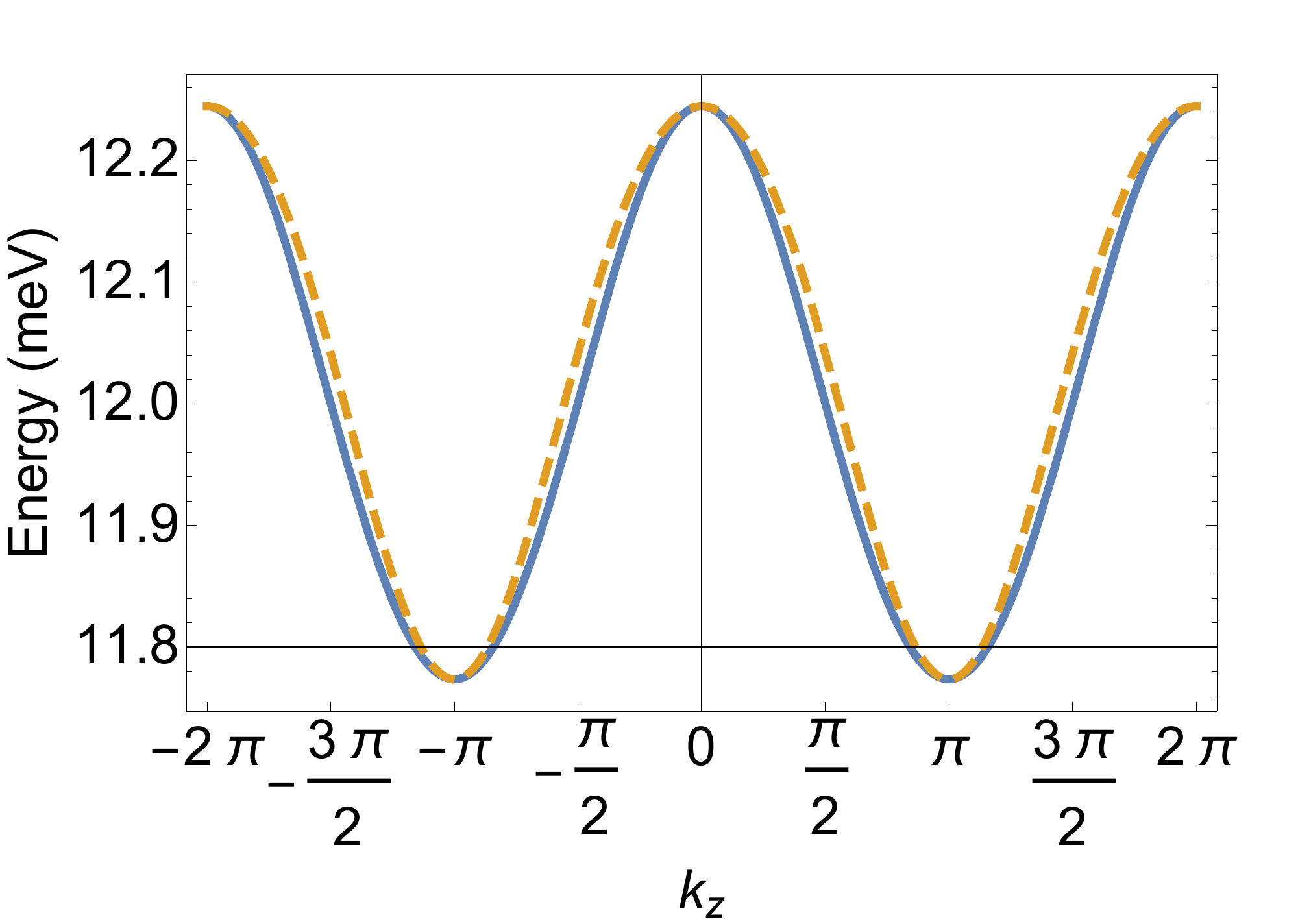}
	}
	\subfloat[Component phases of the opticlike magnon state \label{subfig:exact_vs_approximated_5}]{
		\includegraphics[width=0.33 \textwidth]{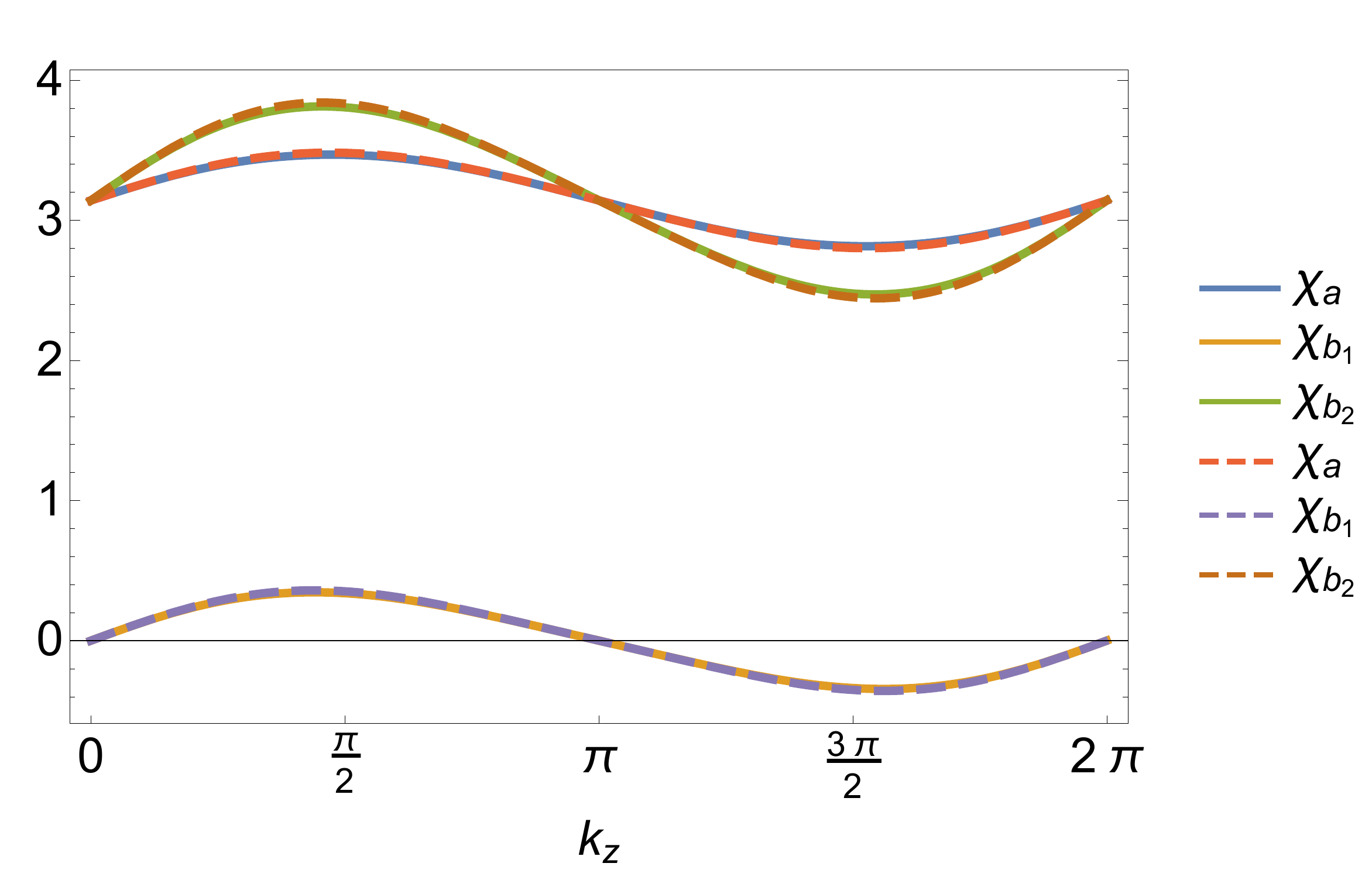}
	}
	\subfloat[Amplitudes on each component of the opticlike magnon state \label{subfig:exact_vs_approximated_6}]{
		\includegraphics[width=0.33 \textwidth]{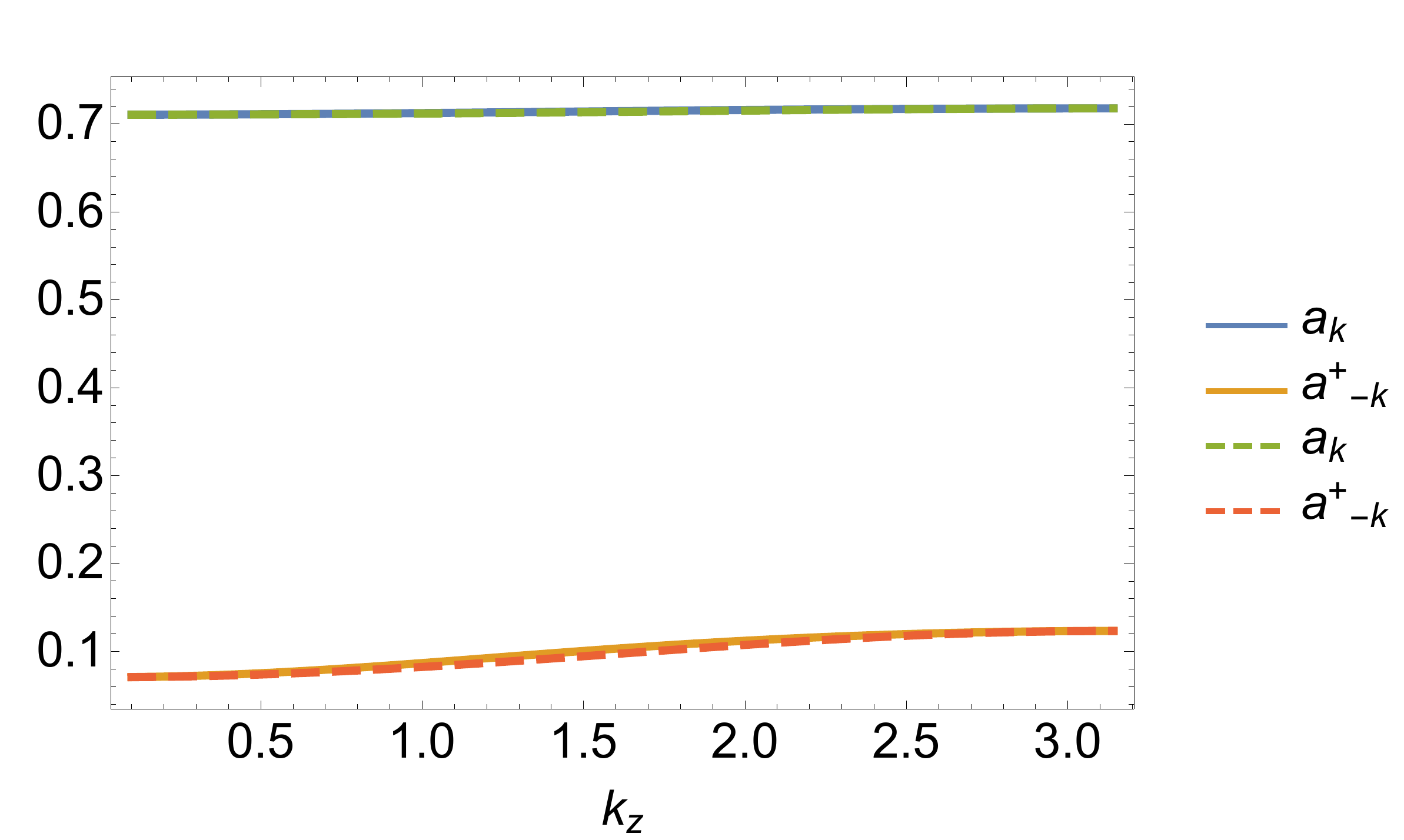}
	}\\
	
	\protect\caption{The acousticlike (first row) and opticlike (second row) branches of the magnon spectrum found by the Holstein-Primakoff method. Plot (a) and (d) give the spectrum; (b) and (e) are phases, while (c) and (f) are magnitudes of each of the components. Here, we plotted the dependence on $k_z$ with $k_x=k_y=0$. The solid curves represent the exact solutions for the eigenvalue equations$(H_{1k}+H_{2k})\psi_1=E_k\sigma_3\psi_1$, while the dashed curves stand for the approximated solutions Eqs. (\ref{solution_to_eigenstates_1}) and (\ref{solution_to_eigenstates_2}).}
	
	\label{fig:exact_vs_approximated}
	
\end{figure*}

As for the higher energy state, we use the same ansatz and repeat the above procedures, but expand $\chi_{a}$ and $\chi_{b_2}$ around $\pi$. Eventually we get the results similar to Eqs. (\ref{solution_to_eigenstates_1}) and (\ref{solution_to_eigenstates_2}) but with the following changes: (i) $B_k\rightarrow-B_k$; (ii) $C_k\rightarrow-C_k$; and finally (iii) phase $\pi$ has to be added to $\chi_{a}$ and $\chi_{b_2}$.

In Fig. \ref{fig:exact_vs_approximated}, we compare the results obtained for the acousticlike and opticlike branches of the magnon spectrum by solving the eigenvalue equation exactly, and using the approximate equations (\ref{solution_to_eigenstates_1}) and (\ref{solution_to_eigenstates_2}). In all six plots, the dashed curves (approximate) almost match the solid ones (exact). Hence, the approximate Eqs. (\ref{solution_to_eigenstates_1}) and (\ref{solution_to_eigenstates_2}) work perfectly.

\section{Results and Discussion}\label{results_and_discussion}
\begin{figure}[htp] \centerline{\includegraphics[clip, width=1  \columnwidth]{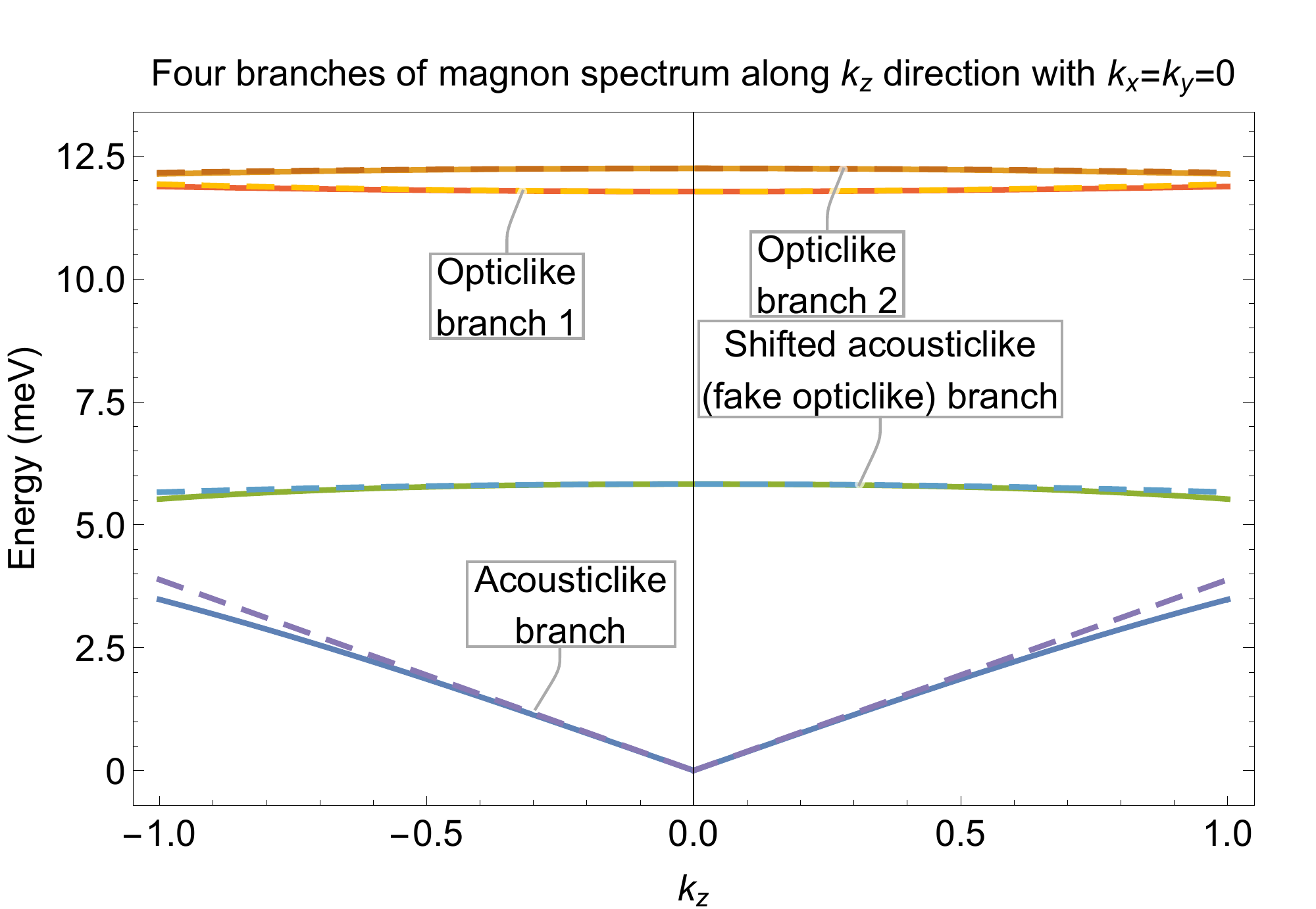}}
	
	\protect\caption{Four branches of the magnon spectrum obtained by the Holstein-Primakoff approach (solid curves) and macroscopic description (dashed curves). The eigenvalues are plotted along the $k_z$ direction with $k_x=k_y=0$.}
	
	\label{fig:four_branches_comparison}
	
\end{figure}
\begin{figure*}[htp]
	\subfloat[Acousticlike branch along $k_z$ direction \label{subfig:exact_vs_macroscopic_a_kz}]{
		\includegraphics[width=0.24 \textwidth]{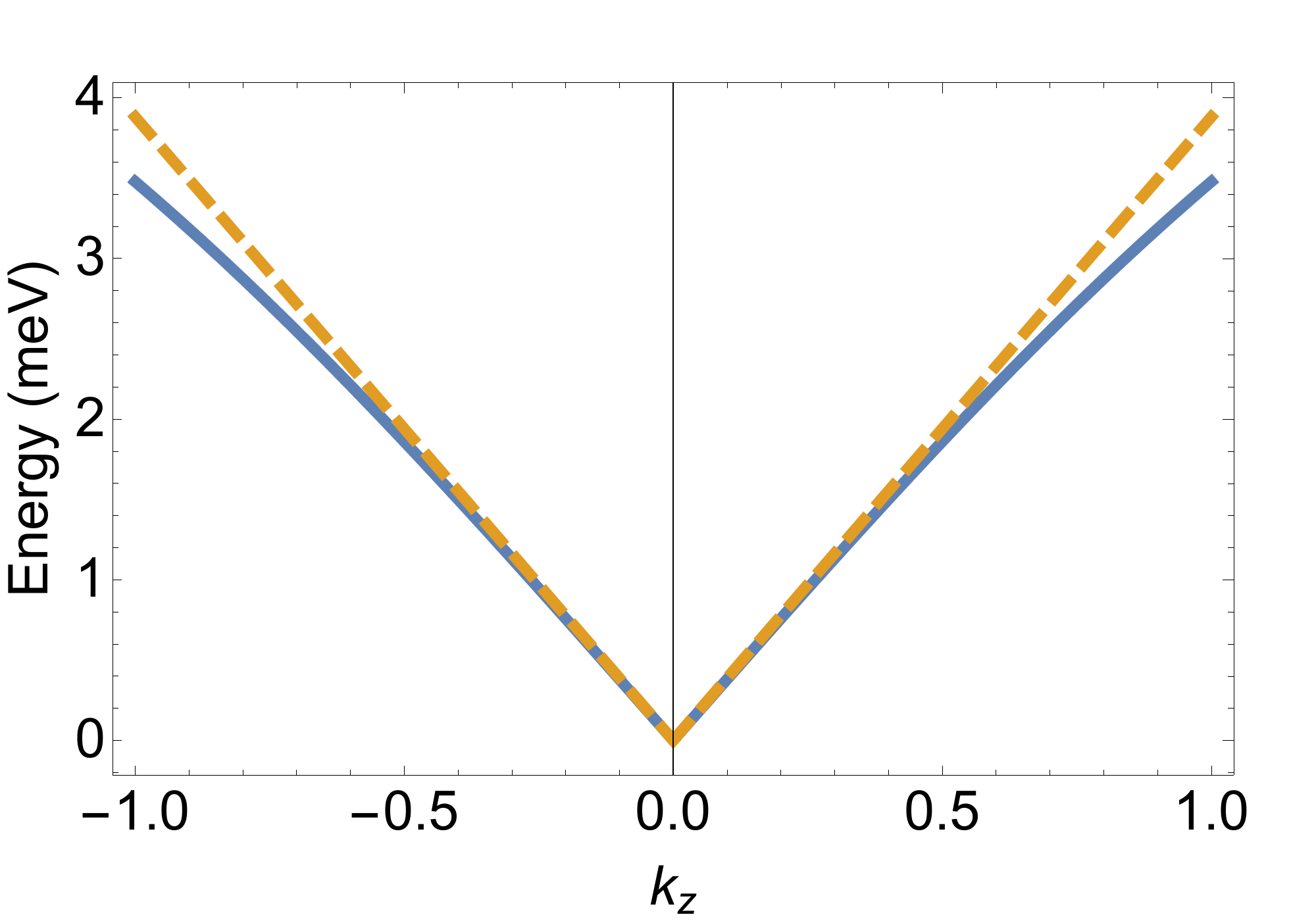}
	}
	\subfloat[Shifted acousticlike branch along $k_z$ direction \label{subfig:exact_vs_macroscopic_o1_kz}]{
		\includegraphics[width=0.24 \textwidth]{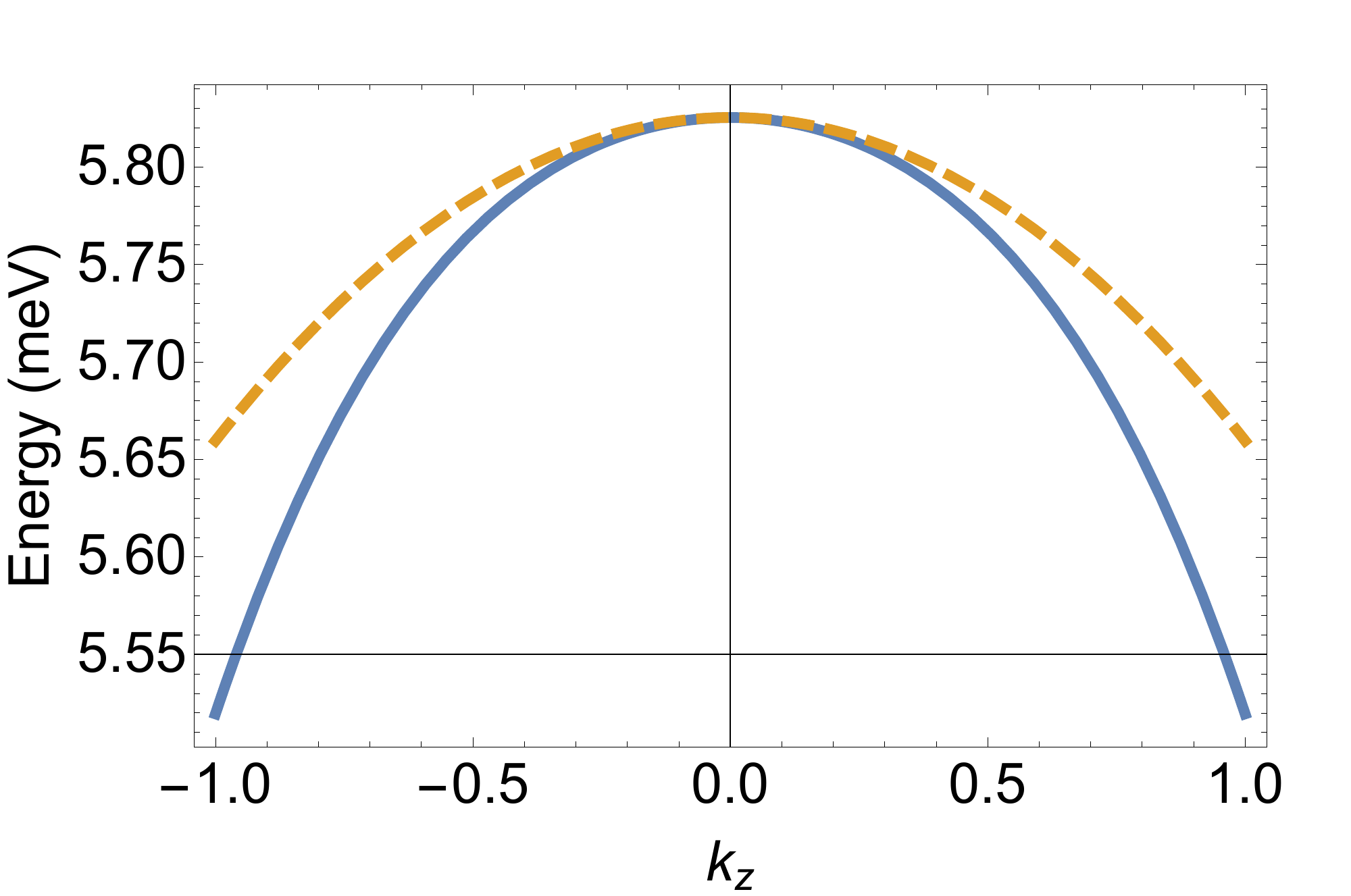}
	}
	\subfloat[Opticlike branch 1 along $k_z$ direction \label{subfig:exact_vs_macroscopic_o2_kz}]{
		\includegraphics[width=0.24 \textwidth]{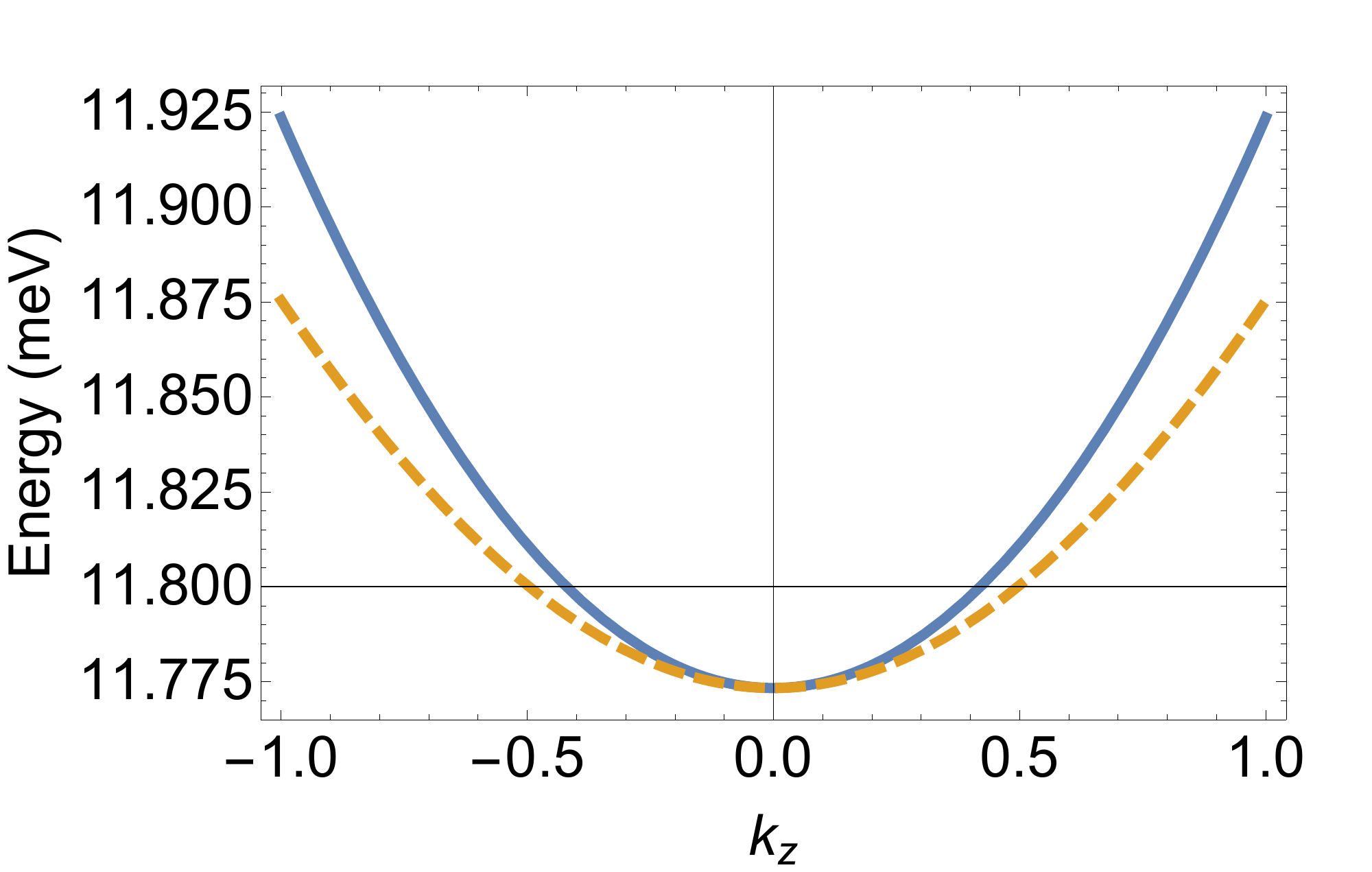}
	}
	\subfloat[Opticlike branch 2 along $k_z$ direction \label{subfig:exact_vs_macroscopic_o3_kz}]{
		\includegraphics[width=0.24 \textwidth]{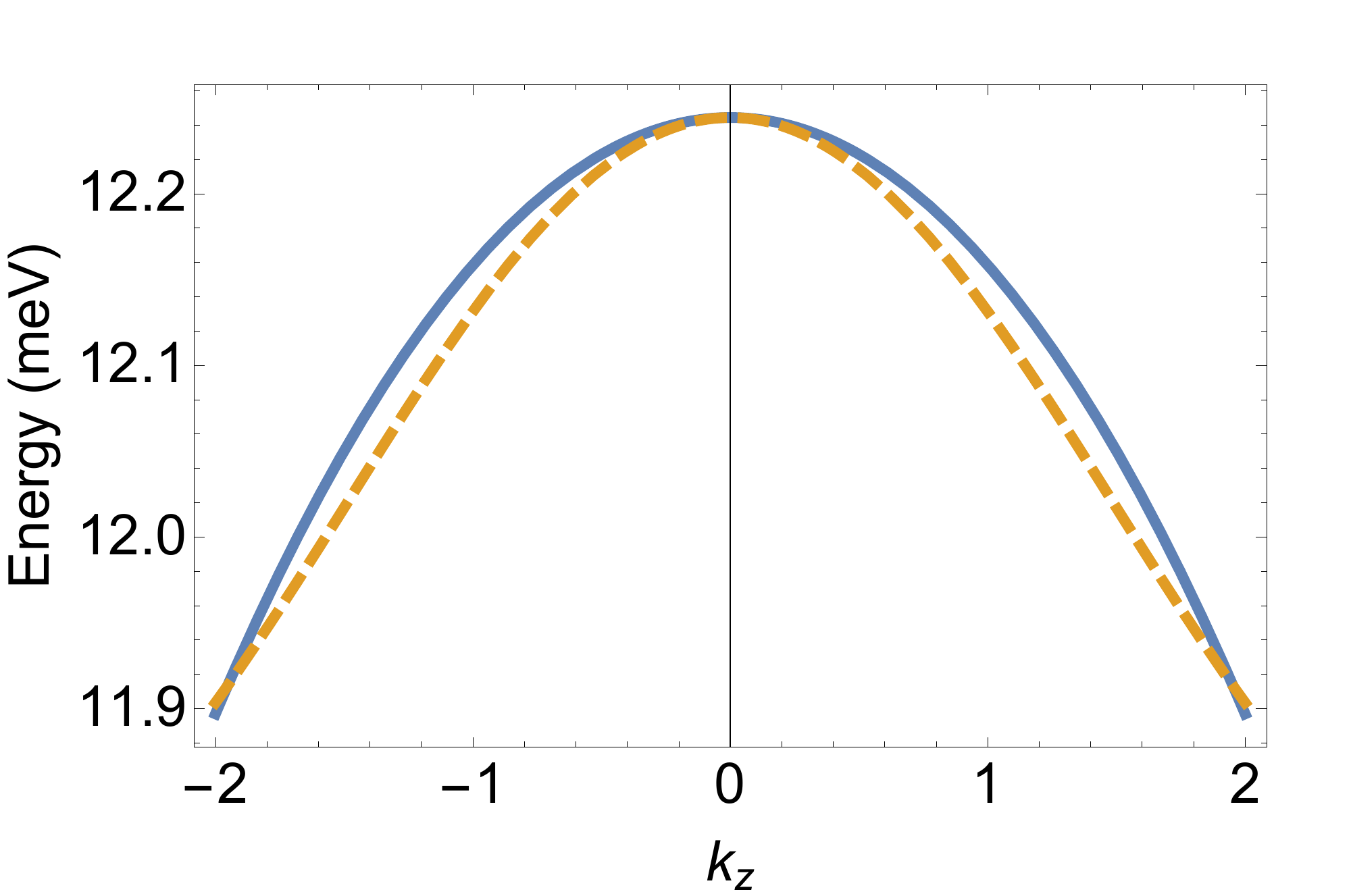}
	}\\
	
	\subfloat[Acousticlike branch along $k_x$ direction \label{subfig:exact_vs_macroscopic_a_kx}]{
		\includegraphics[width=0.24 \textwidth]{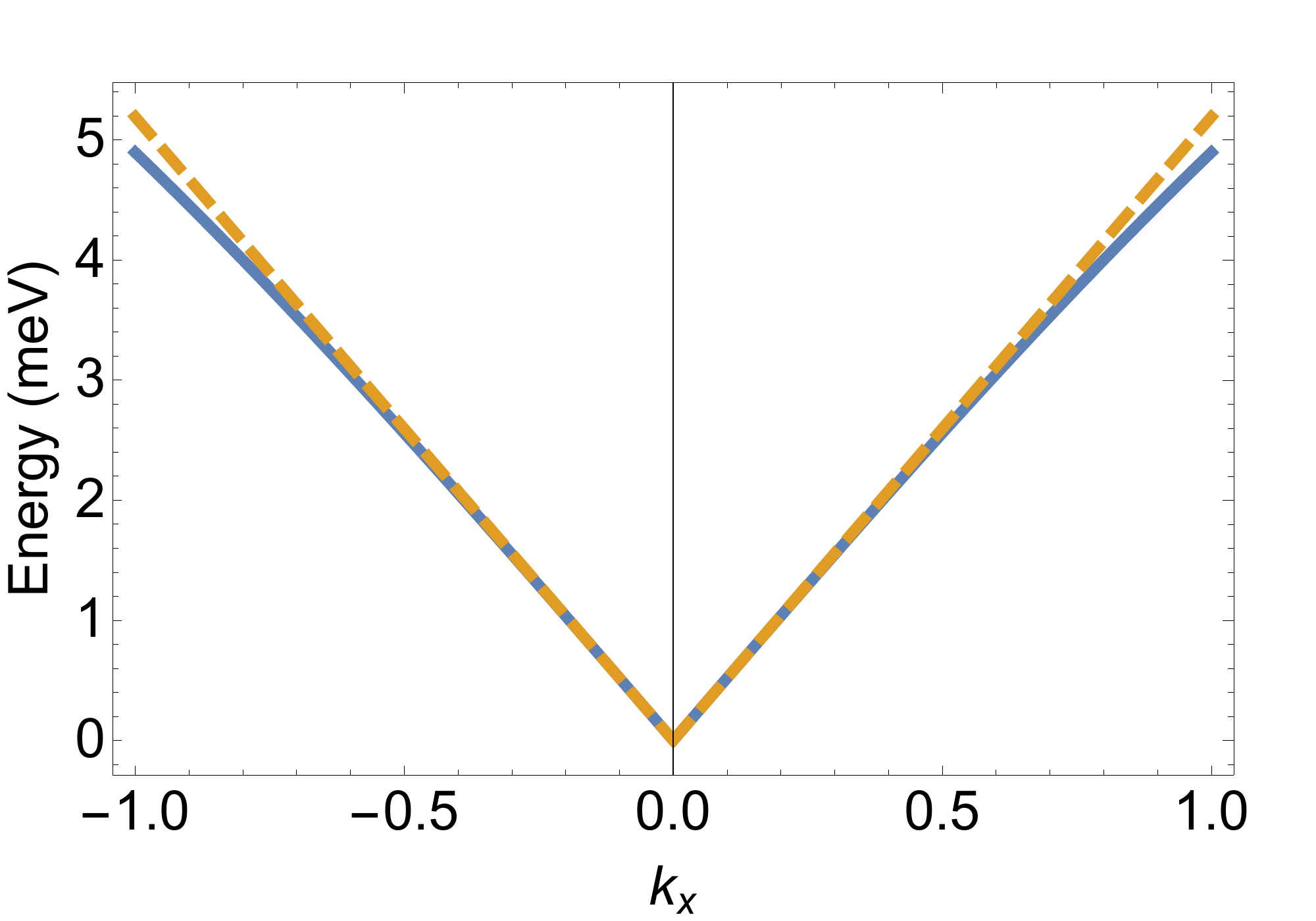}
	}
	\subfloat[Shifted acousticlike branch along $k_x$ direction \label{subfig:exact_vs_macroscopic_o1_kx}]{
		\includegraphics[width=0.24 \textwidth]{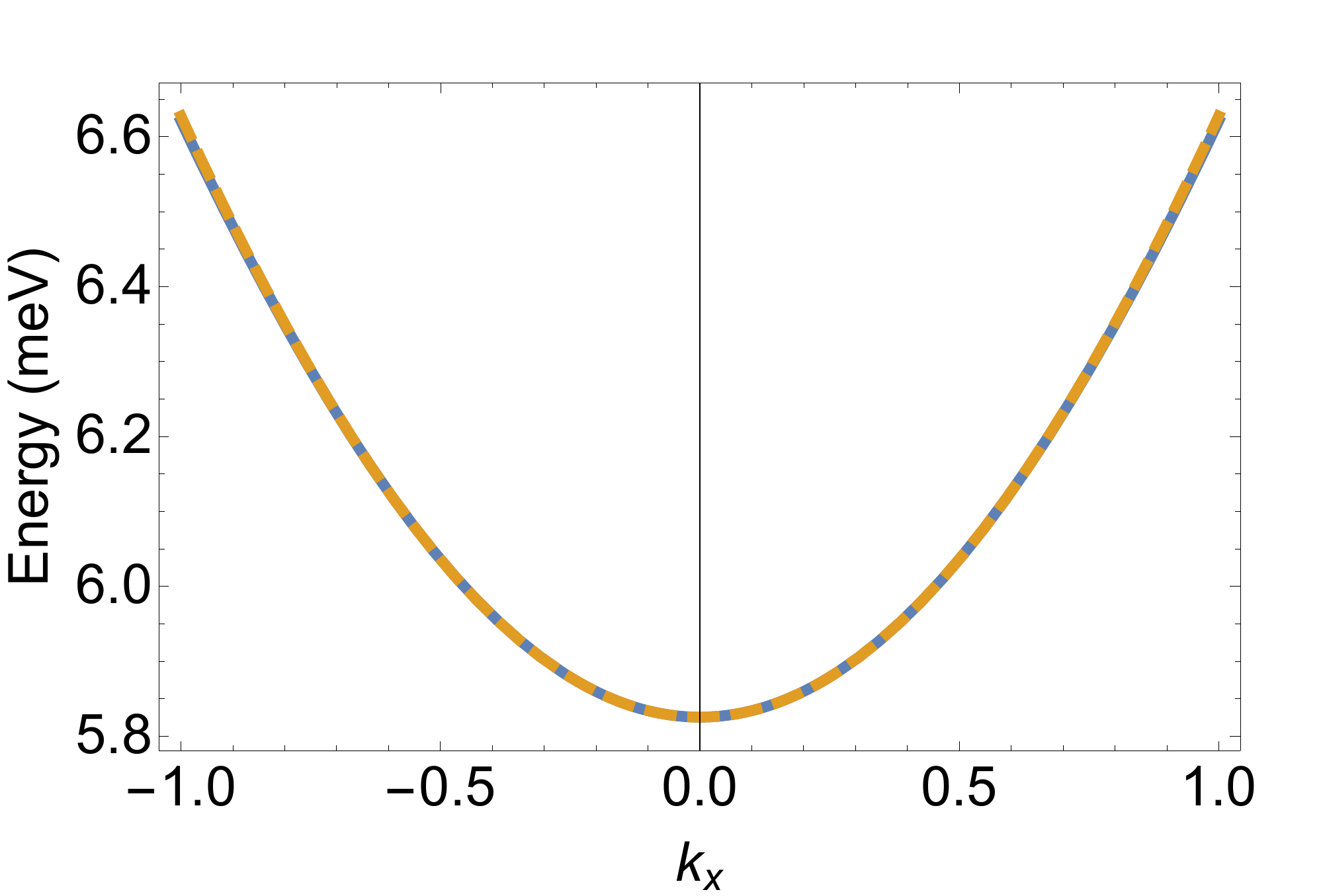}
	}
	\subfloat[Opticlike branch 1 along $k_x$ direction \label{subfig:exact_vs_macroscopic_o2_kx}]{
		\includegraphics[width=0.24 \textwidth]{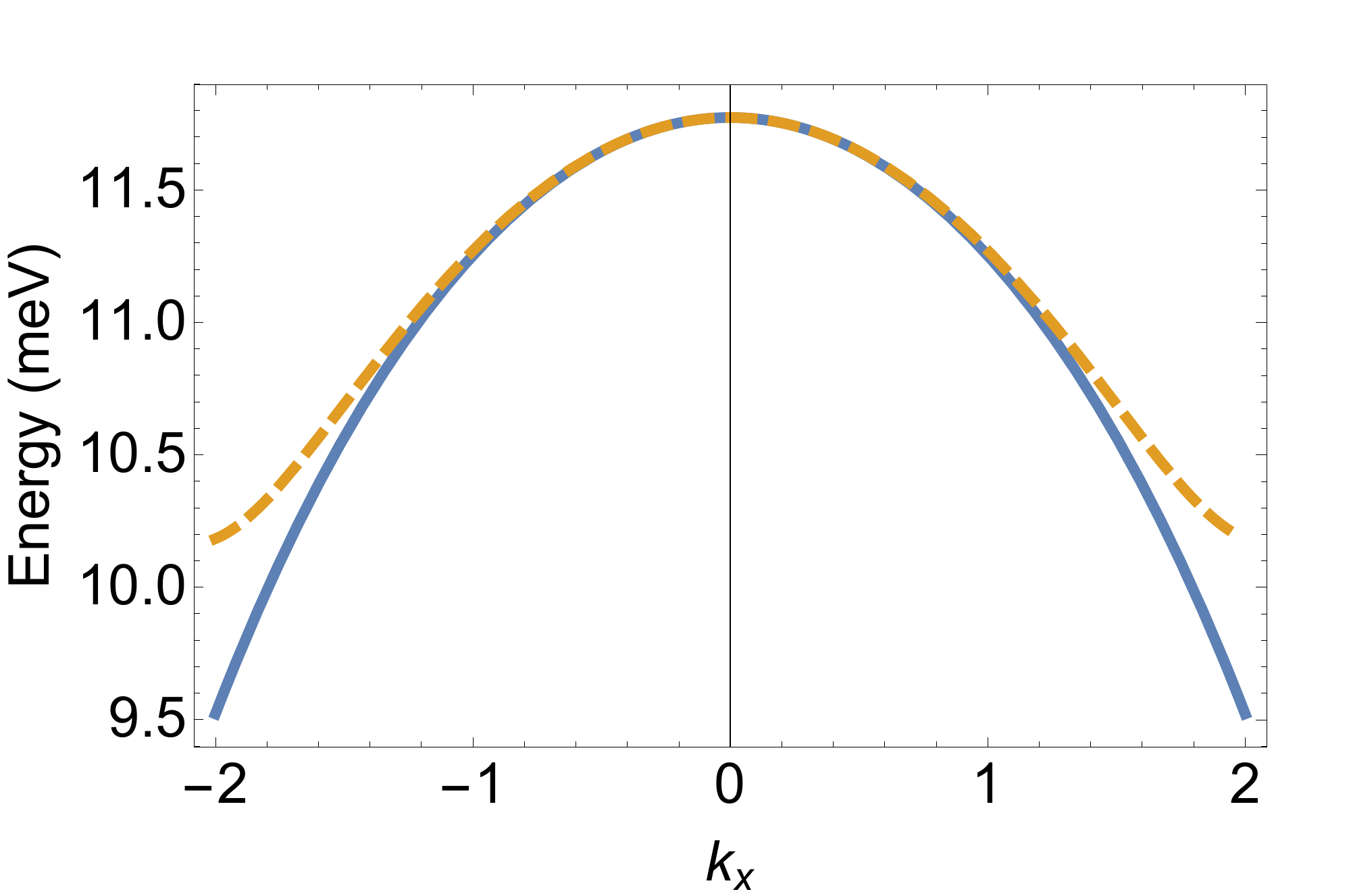}
	}
	\subfloat[Opticlike branch 2 along $k_x$ direction \label{subfig:exact_vs_macroscopic_o3_kx}]{
		\includegraphics[width=0.24 \textwidth]{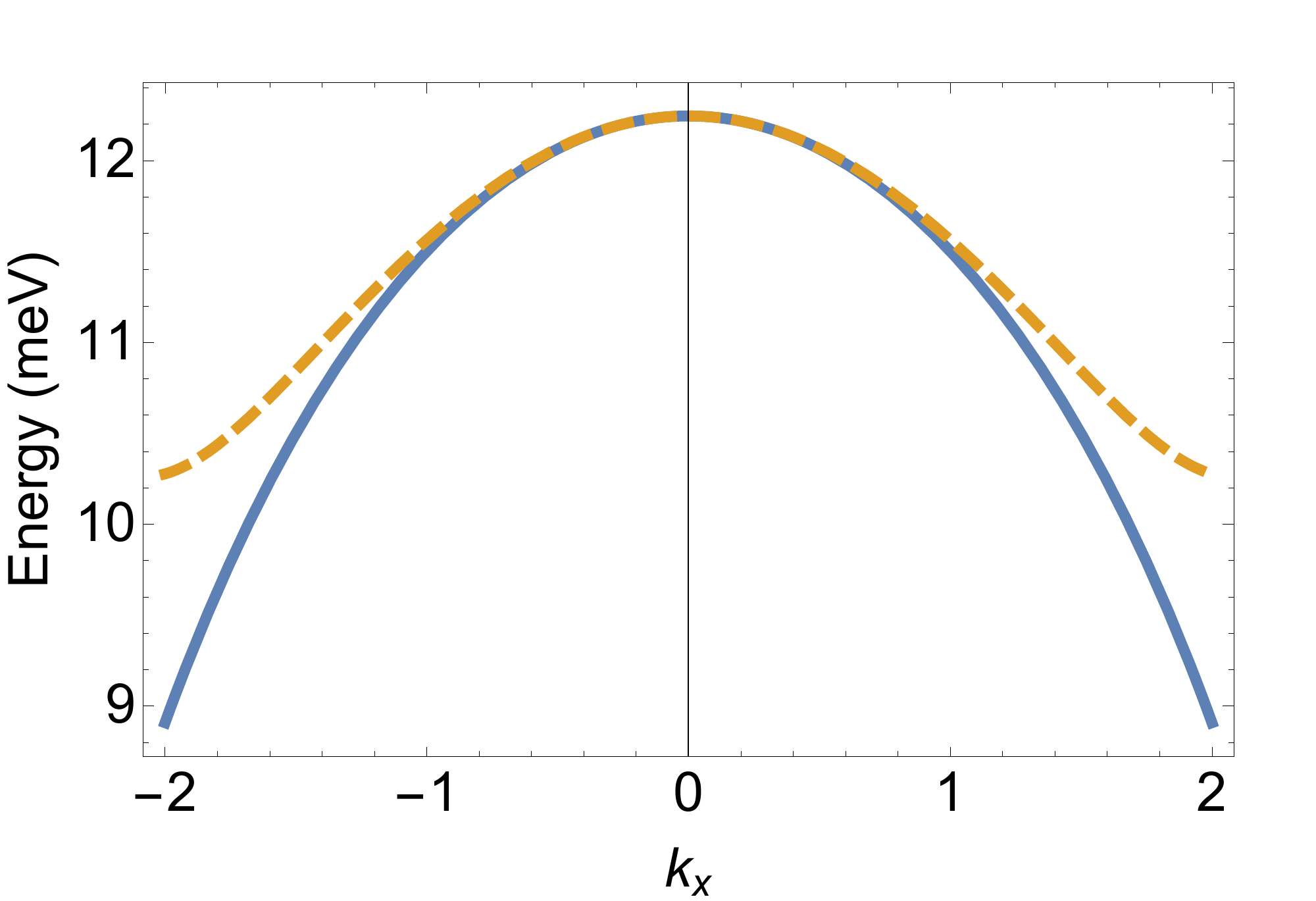}
	}\\
	
	\protect\caption{The fragments of the spectrum of the acousticlike, shifted and two opticlike magnon modes. The first row presents the spectrum along the $k_z$ direction with $k_x=k_y=0$, while the second row gives the $k_x$ direction with $k_y=k_z=0$. (a) and (e) are acousticlike branch; (b) and (f) are the fragments of the shifted acousticlike branch, while (c), (g), and (d), (h) are related to the opticlike branches 1, and 2, respectively. In each subfigure, the solid curve represents the exact solutions for the eigenvalue equations $(H_{1k}\pm H_{2k})\psi_1=E_k\sigma_3\psi_1$, while the dashed curve is described by the semimacroscopic equations (\ref{eigen_values}) and (\ref{eigen_values_higher}).}
	
	\label{fig:exact_vs_macroscopic}
\end{figure*}
Let us compare the results of the Holstein-Primakoff approach with those obtained using the magnetization and N\'eel vector densities. To give a general picture, in Fig. \ref{fig:four_branches_comparison}, we plot four branches of the magnon spectrum given by two methods. As it is shown, each pair of two branches obtained through two different approaches is approximately matching with each other at small $k_z$.

\subsection{Comparison of different methods}\label{comparison}
We now compare, in detail, the results of our description in terms of the macroscopic variables with those obtained from the exact solution of the $8\times8$ model. In Fig. \ref{fig:exact_vs_macroscopic}, we plot the dispersion of the acousticlike and opticlike modes along the $k_z$ and $k_x$ directions. To quantitatively compare the results obtained by these two very different approaches, we estimate the spin wave velocities of the acousticlike branch along the $x$ and $z$ directions; see Figs. \ref{fig:exact_vs_macroscopic}(a) and \ref{fig:exact_vs_macroscopic}(e). We find $\frac{\partial\omega_a(\bm{k})}{\partial k_z}|_{\bm{k}=0}\approx3.81$ for the $8\times8$ model and $\frac{\partial\omega_a(\bm{k})}{\partial k_z}|_{\bm{k}=0}\approx3.88$ for our proposed macroscopic description. The mismatch is less than $2\%$. Along $x$ direction, we get an even better agreement with $\frac{\partial\omega_a(\bm{k})}{\partial k_x}|_{\bm{k}=0}$ estimated to be $5.19$ for both models. Consequently, we conclude that our macroscopic description of the acousticlike magnon branch agrees quantitatively well with the exact spectrum under the long wavelength limit, i.e., $k\lesssim0.5$. 

For completeness, we also compare the results obtained for other three branches of the spin waves (see Fig. \ref{fig:four_branches_comparison}). For the shifted acousticlike and two opticlike branches [see Figs. \ref{fig:exact_vs_macroscopic}(b), \ref{fig:exact_vs_macroscopic}(c), and \ref{fig:exact_vs_macroscopic}(d)], using the Holstein-Primakoff method, we obtain for the $k_z$ direction $\frac{\partial^2\omega_{sh}(\bm{k})}{\partial^2 k_z}|_{\bm{k}=0}\approx-0.41$, $\frac{\partial^2\omega_{o1}(\bm{k})}{\partial^2 k_z}|_{\bm{k}=0}\approx0.22$, and $\frac{\partial^2\omega_{o2}(\bm{k})}{\partial^2 k_z}|_{\bm{k}=0}\approx-0.24$. At the same time, our semimacroscopic approach yields, $\frac{\partial^2\omega_{sh}(\bm{k})}{\partial^2 k_z}|_{\bm{k}=0}\approx-0.33$, $\frac{\partial^2\omega_{o1}(\bm{k})}{\partial^2 k_z}|_{\bm{k}=0}\approx0.30$, and $\frac{\partial^2\omega_{o2}(\bm{k})}{\partial^2 k_z}|_{\bm{k}=0}\approx-0.17$. On the contrary, for the $k_x$ direction [cf. Figs. \ref{fig:exact_vs_macroscopic}(f), \ref{fig:exact_vs_macroscopic}(g), and \ref{fig:exact_vs_macroscopic}(h)], both methods give the same estimates $\frac{\partial^2\omega_{sh}(\bm{k})}{\partial^2 k_x}|_{\bm{k}=0}\approx1.71$, $\frac{\partial^2\omega_{o1}(\bm{k})}{\partial^2 k_x}|_{\bm{k}=0}\approx-1.02$, and $\frac{\partial^2\omega_{o2}(\bm{k})}{\partial^2 k_x}|_{\bm{k}=0}\approx-1.44$. Although there is a relatively large mismatch between the two approaches for the $k_z$ direction, the dispersion within $(k_x,k_y)$ momentum plane is well captured by our semimacroscopical scheme. We ascribe the discrepancy in the magnon spectrum along the $k_z$ direction to the neglecting of $\pm\frac{\partial}{\partial z}$ terms in Eqs. (\ref{EOM_m_AB_1})--(\ref{EOM_l_AB_3}) when deriving the macroscopic equations of motion, see Appendix \ref{derivation_without_deformation} for the details.

\subsection{Dynamics of four branches in terms of the macroscopic variables} \label{spin_waves}
\begin{figure*}[htp]
	\subfloat[Dynamics of $(m_{\theta},\phi)$ pair around $\bm{k}=0$ \label{subfig:spin_dynamics_1}]{
		\includegraphics[width=0.24 \textwidth]{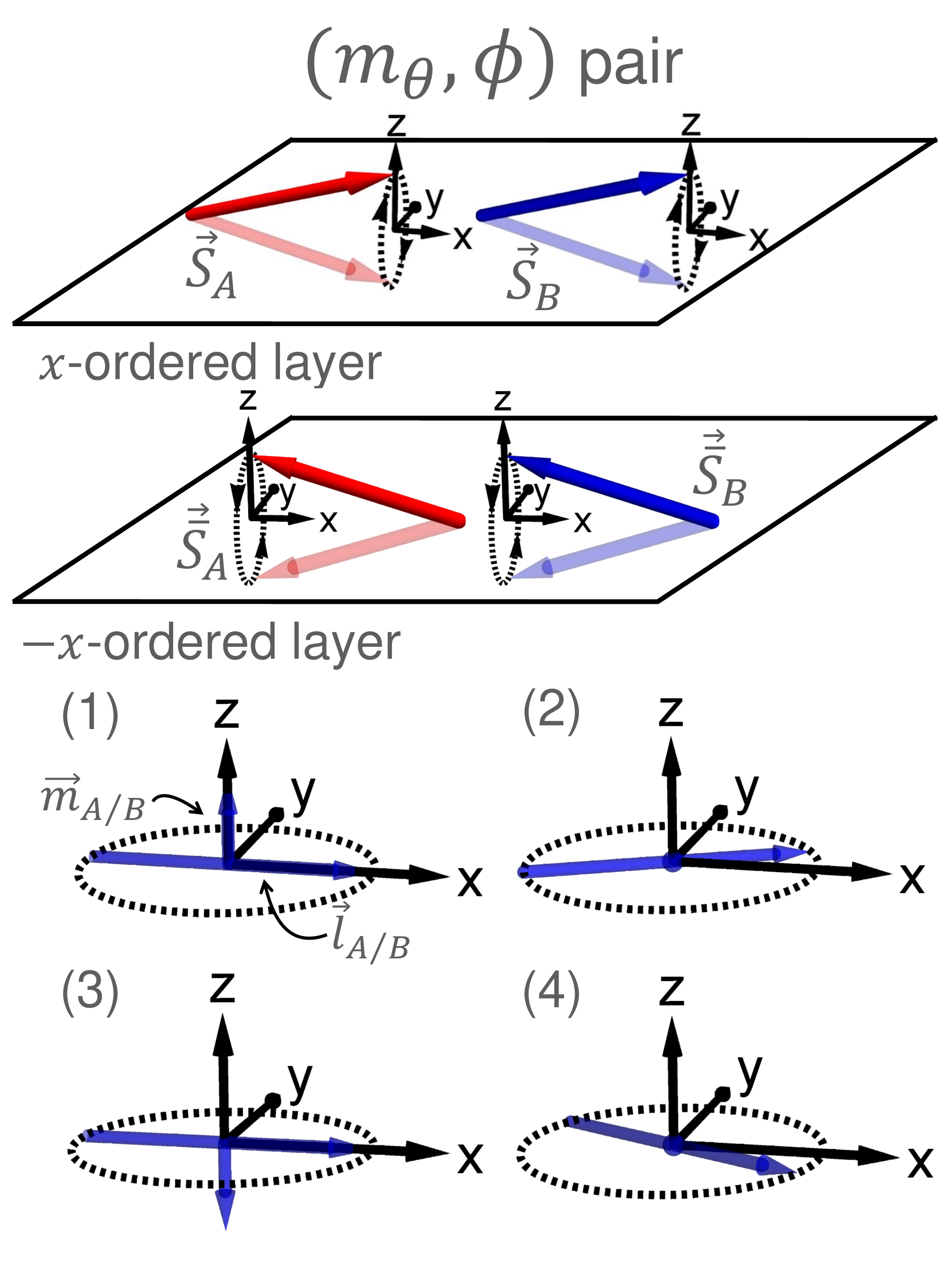}
	}
	\subfloat[Dynamics of $(m_{\phi},\theta)$ pair around $\bm{k}=0$ \label{subfig:spin_dynamics_2}]{
		\includegraphics[width=0.24 \textwidth]{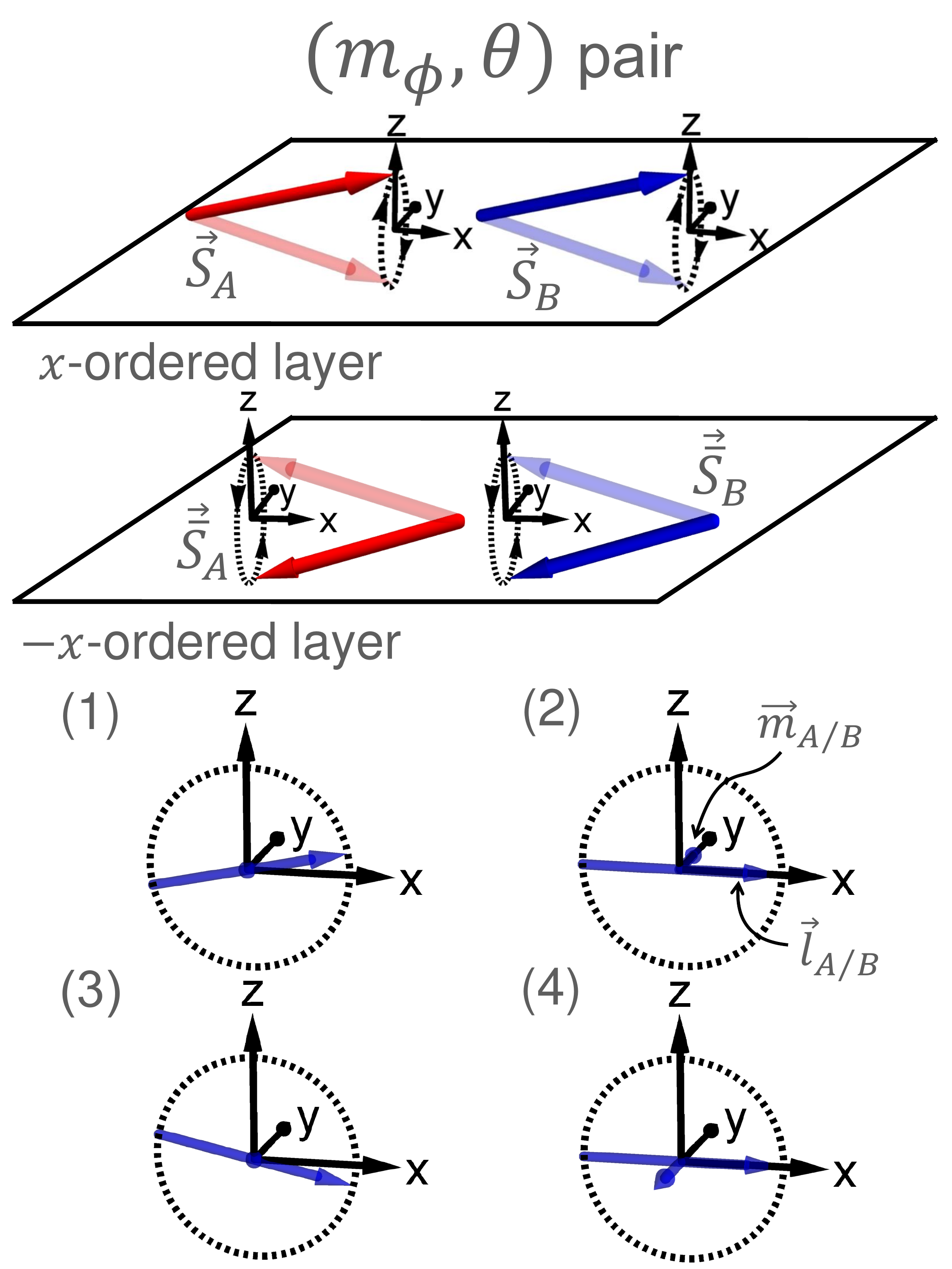}
	}
	\subfloat[Dynamics of $(\delta m^y,\delta l^z)$ pair around $\bm{k}=0$ \label{subfig:spin_dynamics_3}]{
		\includegraphics[width=0.24 \textwidth]{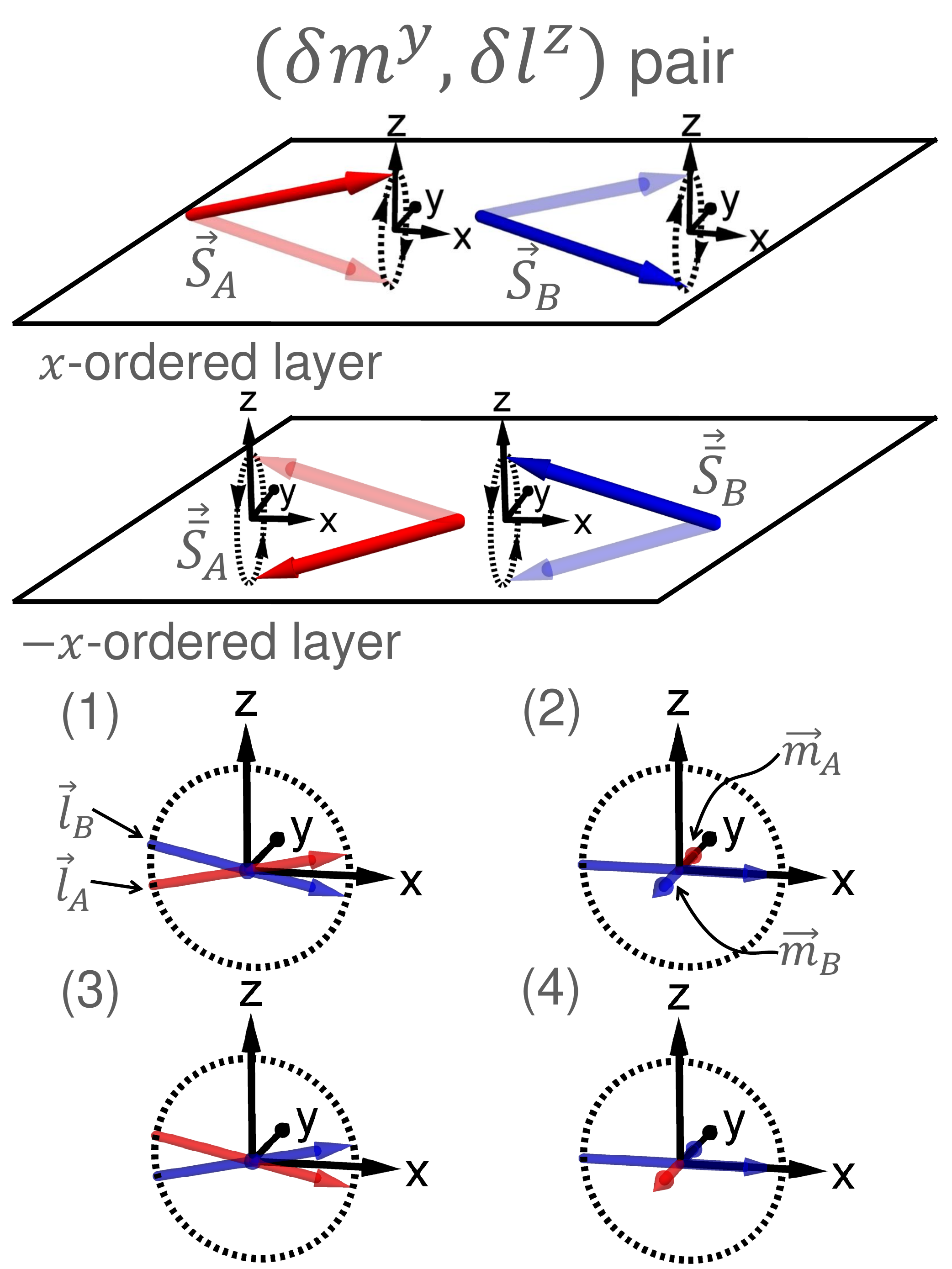}
	}
	\subfloat[Dynamics of $(\delta m^{z},\delta l^{y})$ pair around $\bm{k}=0$ \label{subfig:spin_dynamics_4}]{
		\includegraphics[width=0.24 \textwidth]{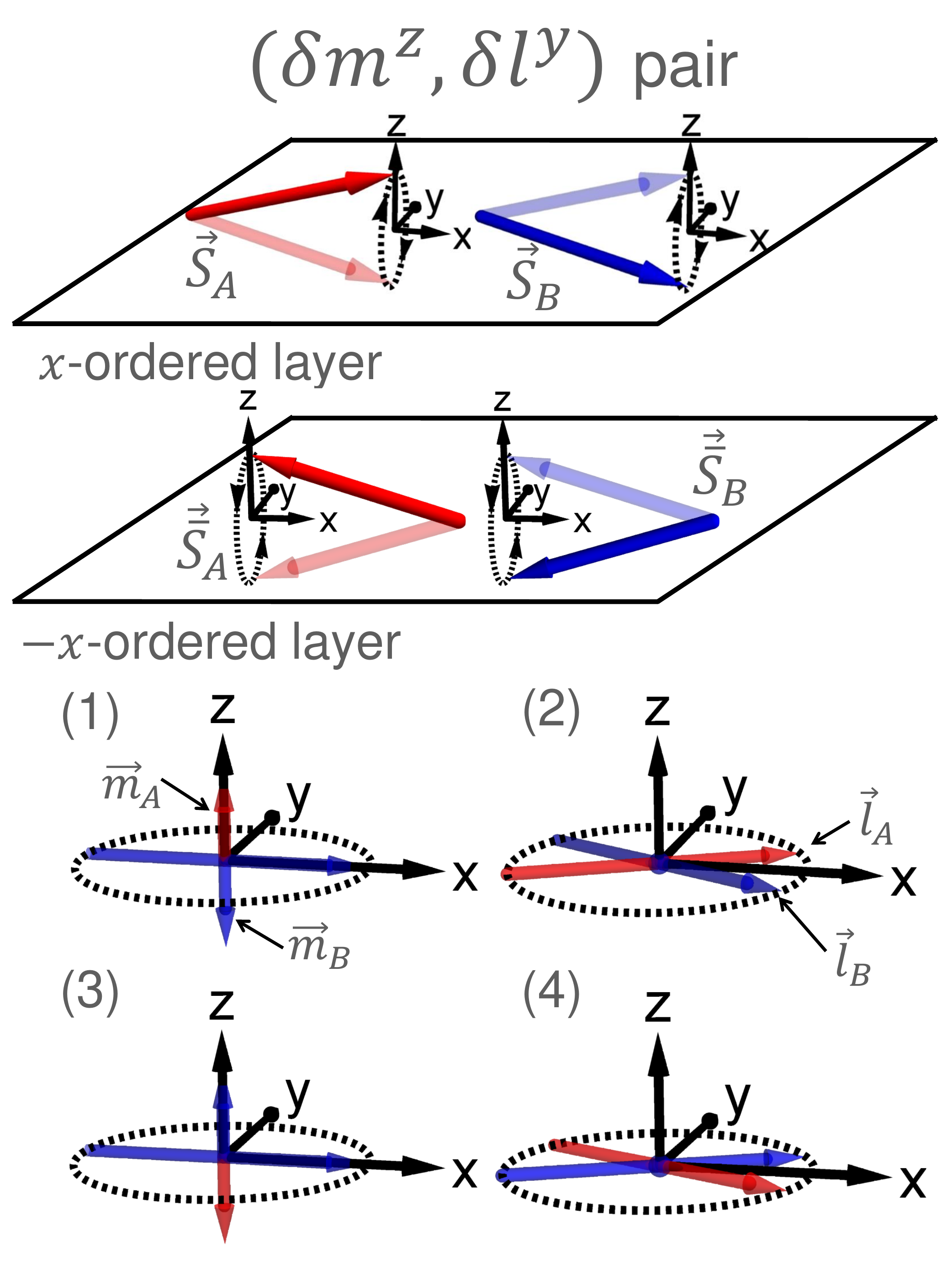}
	}\\
	
	\protect\caption{Dynamics of the four spin wave excitations in terms of the macroscopic pairs: (a) $(m_{\theta},\phi)$; (b) $(m_{\phi},\theta)$; (c) $(\delta m^y,\delta l^z)$; and (d) $(\delta m^{z},\delta l^{y})$ at $\bm{k}\approx0$. Spin vectors, magnetization densities, and N\'eel vectors on the $A$ and $B$ sublattices are indicated by red and blue colors, respectively. In the lower part of each subfigure, (1), (2), (3), and (4) illustrate the magnitudes and directions of $\bm{m}_{A/B}$ and $\bm{l}_{A/B}$ at $t=0$, $\frac{T}{4}$, $\frac{T}{2}$, and $\frac{3T}{4}$, respectively; $T$ is the period of the spin wave.}
	
	\label{fig:spin_dynamics}
\end{figure*}
In this section, to have a better understanding of the four spin-wave branches shown in Fig. \ref{fig:four_branches_comparison}, we give the schematic pictures of their spin dynamics. We start with the acousticlike magnon branch, i.e., the $(m_{\theta},\phi)$ pair. As it is shown in Fig. \ref{fig:spin_dynamics}(a), spins on $A$ and $B$ sublattices are fully synchronized. The magnetization densities $\bm{m}_{A/B}$ alternate along the $z$ direction, while the N\'eel vectors $\bm{l}_{A/B}$ rotate back and forth around the equilibrium position within the $xy$ plane. The magnon frequency of this mode goes to zero as $k\rightarrow0$, because of the rotational symmetry. Generally, this mode manifests the possibility of the spin superfluidity in a system with $XY$ symmetry, see, e.g., Refs. \cite{takei2014superfluid} and \cite{sonin2010spin}. In contrast to the acousticlike branch, the other branch described by the $(m_{\phi}, \theta)$ pair is looking like an opticlike branch due to its finite energy at $k=0$ but, in fact, is the \emph{shifted} version of the acousticlike branch. This mode has the same dynamics on both $A$ and $B$ sublattices. It exhibits alternating $\bm{m}_{A/B}$ along the $y$ direction and rotating $\bm{l}_{A/B}$ within the $xz$ plane. Because of absence of the rotational symmetry around the $y$ direction, this mode gets a finite energy at $k=0$. Note, however, that at $k_z=\pi$, the phase difference between two neighboring layers will interchange the picture of spin dynamics presented in Figs. \ref{fig:spin_dynamics}(a) and \ref{fig:spin_dynamics}(b). At $k_z=\pi$, the discussed branch (looking like the optic one) touches zero, while the acousticlike magnon acquires the finite frequency. This is the reason why we named this branch as the ``shifted acousticlike"; see Fig. \ref{fig:four_branches}.

The dynamics of the other two (true) opticlike branches with finite energies along the whole spectrum are depicted in Figs. \ref{fig:spin_dynamics}(c) and \ref{fig:spin_dynamics}(d). The opticlike branch 1 is similar to the shifted acousticlike branch with $\bm{m}_{A/B}$ alternating along the $y$ direction and $\bm{l}_{A/B}$ rotating within the $xz$ plane. The decisive point here is that spins on $A$ and $B$ sublattices change oppositely. Finally, the opticlike branch 2 is a gapped analog of acousticlike branch: it has out-of-layer magnetization densities and in-layer N\'eel vectors. However, spins on sites $A$ and $B$ evolve oppositely, which makes this mode to be opticlike.

In Appendix \ref{XY_model}, we evaluated the parameters of the Hamiltonian (\ref{H_spins_explicit}) using results obtained here for the four spin-wave modes.

\section{Spin Dynamics in the Presence of Lattice Deformations}\label{macro_and_deformed}
The successful description of the acousticlike magnon excitations by the two different methods encourages us to extend the scheme developed in Sec. \ref{macro_and_free} to a system with a deformation of the lattice. Lattice deformations change the equations of motion Eqs. (\ref{EOM_m_theta}) and (\ref{EOM_m_phi}) obtained in Sec. \ref{macro_and_free}. The point is that deformations change distances between spins that in turn modify exchange coupling constants $J_{\parallel}$ and $J_{\perp}$ in Eq. (\ref{EOM_S}). The changes of $J_{\parallel}$ and $J_{\perp}$ along $\bm{\delta}_1$ and $\bm{\delta}_2$ directions, denoted accordingly as $\delta J_{\parallel}^{\delta_1}$ and $\delta J_{\perp}^{\delta_2}$, are connected with the deformation as follows: $\delta J_{\parallel}^{\delta_1}\sim\frac{1}{a}(\frac{\partial J_{\parallel}}{\partial a})\bm{\delta}_{1}\cdot(\bm{\delta}_{1}\cdot\bm{\nabla}\bm{u})$ and $\delta J_{\perp}^{\delta_2}\sim\frac{1}{c}(\frac{\partial J_{\perp}}{\partial c})\bm{\delta}_{2}\cdot(\bm{\delta}_{2}\cdot\bm{\nabla}\bm{u})$. Here, $a$ and $c$ are the dimensionless intra- and interlayer distances, and $\bm{u}$ is the lattice displacement. As a result, for the in-plane exchange couplings describing an action on a spin located on the $A$ sublattice by those on the $B$ sublattice, i.e., $B\rightarrow A$, we have
\begin{align}\label{delta_J}
	\delta J_{\parallel}^{\delta_{1,1}}&\sim g_1\epsilon_{xx},
	\nonumber
	\\
	\delta J_{\parallel}^{\delta_{1,2}}&\sim g_1(\frac{1}{4}\epsilon_{xx}+\frac{3}{4}\epsilon_{yy}-\frac{\sqrt{3}}{2}\epsilon_{xy}),
	\nonumber
	\\
	\delta J_{\parallel}^{\delta_{1,3}}&\sim g_1(\frac{1}{4}\epsilon_{xx}+\frac{3}{4}\epsilon_{yy}+\frac{\sqrt{3}}{2}\epsilon_{xy}).
\end{align}
Here, $g_1\equiv\frac{1}{a}(\frac{\partial J_{\parallel}}{\partial a})$, and the strain tensor $\epsilon_{\alpha\beta}\equiv\frac{1}{2}(\partial_{\alpha}u^{\beta}+\partial_{\beta}u^{\alpha})$ with $\alpha,\beta=x,y,z$. For the interlayer exchange couplings, this idea works similarly, and finally, with the use of the standard parametrization, we find that in the presence of a lattice deformation the linearized equations for $m_{\theta}$, $m_{\phi}$, $\theta$, and $\phi$ become [c.f.  Eqs. (20) and (21), the comprehensive derivation is shown in Appendix \ref{derivation_with_deformation}]
\begin{align}\label{EOM_m_theta_delta_J}
	&\dot{m}_{\theta}\approx(4\tilde{S}^{2})(-\frac{3}{8}J_{\parallel}\nabla_{-}^{2}\phi+\frac{9}{8}J_{\perp}\nabla_{+}^{2}\phi),
	\nonumber
	\\
	&\dot{\phi}\approx(-\frac{3}{2}\tilde{J}_{\parallel}+9\tilde{J}_{\perp})m_{\theta};
\end{align}
and
\begin{align}\label{EOM_m_phi_delta_J}
	&\dot{m}_{\phi}\approx(4\tilde{S}^{2})(-\frac{3}{2}\tilde{J}_{\parallel}\theta-\frac{9}{8}J_{\perp}\nabla_{+}^{2}\theta),
	\nonumber
	\\
	&\dot{\theta}\approx(-9\tilde{J}_{\perp}-\frac{3}{8}J_{\parallel}\nabla_{-}^{2}-\frac{9}{8}J_{\perp}\nabla_{+}^{2})m_{\phi}.
\end{align}
Here, exchange coefficients are modified by the strain tensor
\begin{align}
\tilde{J}_{\parallel}\equiv J_{\parallel}+\frac{1}{2}g_1(\epsilon_{xx}+\epsilon_{yy})
\end{align} 
and 
\begin{align}
\tilde{J}_{\perp}\equiv J_{\perp}+\frac{1}{2}g_2(\epsilon_{xx}+\epsilon_{yy}+2\epsilon_{zz})
\end{align}
with $g_2\equiv\frac{1}{c}\frac{\partial J_{\perp}}{\partial c}$ to be the out-of-plane magnetoelastic coefficient. The equations (\ref{EOM_m_theta_delta_J}) and (\ref{EOM_m_phi_delta_J}) are one of the main results of this work. In Ref. [\onlinecite{liu2022control}] we used these equations for description of scattering of the AFM magnons in the backward direction.

\section{Concluding Remarks}\label{Conclusion}
In this paper, we studied the dynamics of spins in a layered van der Waals crystal CoTiO$_3$. This system is a $3D$ quantum $XY$ AFM, with the direction of magnetization alternating between the neighboring layers. As is well known, the $XY$ AFMs are spin analogues of the superfluid Helium and superconductors \cite{hohenberg1977theory,sonin2010spin}. The angle of orientation of the N\'eel vector is equivalent to the superfluid phase. Correspondingly, the long-wavelength magnons are the Goldstone excitations in an $XY$ AFM. We have studied the spectrum of magnons using corresponding pairs of the macroscopic quantities, which are the magnetization and the N\'eel vector densities of various kind. We demonstrate here that for the acousticlike excitations (i.e., for the Goldstone mode), the accuracy of the scheme is almost perfect. Besides, we have confirmed the $XY$ type of the intralayer spin exchange in this material by comparing our macroscopic description of the all four spin-wave modes with the experimental data.

In addition to the spectrum of magnons, we considered the case when the crystal lattice of the magnetic substance is deformed by an external strain. One may expect \cite{liu2022control} that the spin flow could be manipulated by applying a spatially modulated strain. The description of the quantum AFM developed in this paper provides a simple ready-to-use scheme for studying the spin superfluidity in such magnetic systems, as well as the possibility to control the spin dynamics through the lattice deformation.

We would like to emphasize that at the derivation of the equations of motion for the spin-wave excitations, i.e., Eqs. (\ref{EOM_m_1})--(\ref{EOM_l_3}), the route used in this paper is somewhat different from the one in the conventional approach (cf. Refs. [\onlinecite{haldane1983nonlinear}] and [\onlinecite{auerbach2012interacting}]). Conventionally, one starts from the spin Hamiltonian, then constructs the path integral using the spin coherent states and, finally, obtains the Lagrangian density, which can be recognized as the nonlinear sigma model. Eventually, the equations of motion are found by making the variation of the action to be zero. In the present paper, the order of operations was changed. We started with the derivation of the equations of motion for the quantum spin operators directly from the Hamiltonian. Then, these equations were treated in terms of the continuous variables with a nonlinear constraint. By performing this step, we effectively executed the transition to the language of the nonlinear sigma model.

\begin{acknowledgments}
We thank the Referee of our paper \cite{liu2022control} for the suggestion to publish this material in full detail.
\end{acknowledgments}

\begin{widetext}

\appendix

\section{Derivation of equations of motion for $\bm{m}$ and $\bm{l}$ without deformation}\label{derivation_without_deformation}
In this appendix, we show how to derive Eqs. (\ref{EOM_m_1})--(\ref{EOM_l_3}) and (\ref{EOM_m_x})--(\ref{EOM_l_z}) in the main text. Similar to what were discussed in the main text, for the case when site $i$ is on the $B$ sublattice, Eqs. (\ref{gradient_expansion_1}) and (\ref{gradient_expansion_2}) are modified as follows: (i) $A\leftrightarrow B$; (ii) $\frac{\partial}{\partial z}\rightarrow-\frac{\partial}{\partial z}$. Finally, incorporating the gradient expansion terms in Eq. (\ref{EOM_S}), one obtains the equations of motion for the spin components $\bm{S}_{A/B}$:
\begin{align}\label{EOM_S_AB_1}
	\frac{dS_{A/B}^{x}}{dt}\approx&3J_{\parallel}S_{A/B}^{z}S_{B/A}^{y}+3J_{\perp}[S_{A/B}^{z}(\bar{S}_{B/A}^{y}+2\bar{S}_{A/B}^{y})-S_{A/B}^{y}(\bar{S}_{B/A}^{z}+2\bar{S}_{A/B}^{z})]+\frac{3}{4}J_{\parallel}S_{A/B}^{z}\nabla_{-}^{2}S_{B/A}^{y}
	\nonumber
	\\
	&+3J_{\perp}\Big\{S_{A/B}^{z}[\pm\frac{\partial\bar{S}_{B/A}^{y}}{\partial z}+\frac{1}{4}\nabla_{+}^{2}(\bar{S}_{B/A}^{y}+2\bar{S}_{A/B}^{y})]-S_{A/B}^{y}[\pm\frac{\partial\bar{S}_{B/A}^{z}}{\partial z}+\frac{1}{4}\nabla_{+}^{2}(\bar{S}_{B/A}^{z}+2\bar{S}_{A/B}^{z})]\Big\},
\end{align}
\begin{align}\label{EOM_S_AB_2}
	\frac{dS_{A/B}^{y}}{dt}\approx&-3J_{\parallel}S_{A/B}^{z}S_{B/A}^{x}+3J_{\perp}[S_{A/B}^{x}(\bar{S}_{B/A}^{z}+2\bar{S}_{A/B}^{z})-S_{A/B}^{z}(\bar{S}_{B/A}^{x}+2\bar{S}_{A/B}^{x})]-\frac{3}{4}J_{\parallel}S_{A/B}^{z}\nabla_{-}^{2}S_{B/A}^{x}
	\nonumber
	\\
	&+3J_{\perp}\Big\{S_{A/B}^{x}[\pm\frac{\partial\bar{S}_{B/A}^{z}}{\partial z}+\frac{1}{4}\nabla_{+}^{2}(\bar{S}_{B/A}^{z}+2\bar{S}_{A/B}^{z})]-S_{A/B}^{z}[\pm\frac{\partial\bar{S}_{B/A}^{x}}{\partial z}+\frac{1}{4}\nabla_{+}^{2}(\bar{S}_{B/A}^{x}+2\bar{S}_{A/B}^{x})]\Big\},
\end{align}
and
\begin{align}\label{EOM_S_AB_3}
	\frac{dS_{A/B}^{z}}{dt}\approx&3J_{\parallel}(S_{A/B}^{y}S_{B/A}^{x}-S_{A/B}^{x}S_{B/A}^{y})+3J_{\perp}[S_{A/B}^{y}(\bar{S}_{B/A}^{x}+2\bar{S}_{A/B}^{x})-S_{A/B}^{x}(\bar{S}_{B/A}^{y}+2\bar{S}_{A/B}^{y})]
	\nonumber
	\\
	&+\frac{3}{4}J_{\parallel}(S_{A/B}^{y}\nabla_{-}^{2}S_{B/A}^{x}-S_{A/B}^{x}\nabla_{-}^{2}S_{B/A}^{y})+3J_{\perp}\Big\{S_{A/B}^{y}[\pm\frac{\partial\bar{S}_{B/A}^{x}}{\partial z}+\frac{1}{4}\nabla_{+}^{2}(\bar{S}_{B/A}^{x}+2\bar{S}_{A/B}^{x})]
	\nonumber
	\\
	&-S_{A/B}^{x}[\pm\frac{\partial\bar{S}_{B/A}^{y}}{\partial z}+\frac{1}{4}\nabla_{+}^{2}(\bar{S}_{B/A}^{y}+2\bar{S}_{A/B}^{y})]\Big\}.
\end{align}
Here, we eventually dropped the site index $i$ in the spin operators, assuming from now on that $\bm{S}_{A/B}$ are space- and time-dependent variables $\bm{S}_{A/B}(\bm{r},t)$. For the spatial derivatives, we have also introduced a short notation, $\nabla_{\pm}^{2}\equiv\nabla^{2}\pm\frac{\partial^{2}}{\partial z^{2}}$. The equations of motion for $\bar{\bm{S}}_{A/B}$ could be obtained through the exchange $\bm{S}_{A/B}\leftrightarrow\bar{\bm{S}}_{A/B}$ in the above equations.

Next, we define the total magnetization $\bm{m}_{A/B}\equiv\bm{S}_{A/B}+\bar{\bm{S}}_{A/B}$ and the N\'eel vector $\bm{l}_{A/B}\equiv\bm{S}_{A/B}-\bar{\bm{S}}_{A/B}$ for the $A/B$ sublattices, see, e.g., Ref. \cite{auerbach2012interacting}. Note that in a simple N\'eel antiferromagnet the vectors $\bm{m}$ and $\bm{l}$ are orthogonal, $\bm{m}_{A/B}\cdot\bm{l}_{A/B}=0$. In the following part of this paper, $\bm{m}_{A/B}$ and $\bm{l}_{A/B}$ will be considered as classical variables rather than the quantum operators. The resulting equations of motion for $\bm{m}_{A}$ and $\bm{l}_{A}$ are
\begin{align}\label{EOM_m_AB_1}
	\frac{dm_{A}^{x}}{dt}\approx&\frac{3}{2}J_{\parallel}(m_{A}^{z}m_{B}^{y}+\underline{l_{A}^{z}l_{B}^{y}})+\frac{3}{2}J_{\perp}[(m_{A}^{z}m_{B}^{y}-m_{A}^{y}m_{B}^{z})-(\underline{l_{A}^{z}l_{B}^{y}-l_{A}^{y}l_{B}^{z}})]+\frac{3}{8}J_{\parallel}(m_{A}^{z}\nabla_{-}^{2}m_{B}^{y}+l_{A}^{z}\nabla_{-}^{2}l_{B}^{y})
	\nonumber
	\\
	&+\frac{3}{8}J_{\perp}\Big\{[m_{A}^{z}(\nabla_{+}^{2}+4\frac{\partial}{\partial z})m_{B}^{y}-l_{A}^{z}(\nabla_{+}^{2}+4\frac{\partial}{\partial z})l_{B}^{y}+2m_{A}^{z}\nabla_{+}^{2}m_{A}^{y}-2l_{A}^{z}\nabla_{+}^{2}l_{A}^{y}]-[m_{A}^{y}(\nabla_{+}^{2}+4\frac{\partial}{\partial z})m_{B}^{z}
	\nonumber
	\\
	&-\underline{l_{A}^{y}(\nabla_{+}^{2}+4\frac{\partial}{\partial z})l_{B}^{z}}+2m_{A}^{y}\nabla_{+}^{2}m_{A}^{z}-\underline{2l_{A}^{y}\nabla_{+}^{2}l_{A}^{z}}]\Big\},
\end{align}
\begin{align}\label{EOM_m_AB_2}
	\frac{dm_{A}^{y}}{dt}\approx&-\frac{3}{2}J_{\parallel}(m_{A}^{z}m_{B}^{x}+\underline{l_{A}^{z}l_{B}^{x}})+\frac{3}{2}J_{\perp}[(m_{A}^{x}m_{B}^{z}-m_{A}^{z}m_{B}^{x})-(\underline{l_{A}^{x}l_{B}^{z}-l_{A}^{z}l_{B}^{x}})]-\frac{3}{8}J_{\parallel}(m_{A}^{z}\nabla_{-}^{2}m_{B}^{x}+l_{A}^{z}\nabla_{-}^{2}l_{B}^{x})
	\nonumber
	\\
	&+\frac{3}{8}J_{\perp}\Big\{[m_{A}^{x}(\nabla_{+}^{2}+4\frac{\partial}{\partial z})m_{B}^{z}-\underline{l_{A}^{x}(\nabla_{+}^{2}+4\frac{\partial}{\partial z})l_{B}^{z}}+2m_{A}^{x}\nabla_{+}^{2}m_{A}^{z}-\underline{2l_{A}^{x}\nabla_{+}^{2}l_{A}^{z}}]-[m_{A}^{z}(\nabla_{+}^{2}+4\frac{\partial}{\partial z})m_{B}^{x}
	\nonumber
	\\
	&-l_{A}^{z}(\nabla_{+}^{2}+4\frac{\partial}{\partial z})l_{B}^{x}+2m_{A}^{z}\nabla_{+}^{2}m_{A}^{x}-2l_{A}^{z}\nabla_{+}^{2}l_{A}^{x}]\Big\},
\end{align}
\begin{align}\label{EOM_m_AB_3}
	\frac{dm_{A}^{z}}{dt}\approx&\frac{3}{2}J_{\parallel}[(m_{A}^{y}m_{B}^{x}+\underline{l_{A}^{y}l_{B}^{x}})-(m_{A}^{x}m_{B}^{y}+\underline{l_{A}^{x}l_{B}^{y}})]+\frac{3}{2}J_{\perp}[(m_{A}^{y}m_{B}^{x}-m_{A}^{x}m_{B}^{y})-(\underline{l_{A}^{y}l_{B}^{x}-l_{A}^{x}l_{B}^{y}})]+\frac{3}{8}J_{\parallel}[(m_{A}^{y}\nabla_{-}^{2}m_{B}^{x}
	\nonumber
	\\
	&+\underline{l_{A}^{y}\nabla_{-}^{2}l_{B}^{x}})-(m_{A}^{x}\nabla_{-}^{2}m_{B}^{y}+\underline{l_{A}^{x}\nabla_{-}^{2}l_{B}^{y}})]+\frac{3}{8}J_{\perp}\Big\{[m_{A}^{y}(\nabla_{+}^{2}+4\frac{\partial}{\partial z})m_{B}^{x}-\underline{l_{A}^{y}(\nabla_{+}^{2}+4\frac{\partial}{\partial z})l_{B}^{x}}+2m_{A}^{y}\nabla_{+}^{2}m_{A}^{x}
	\nonumber
	\\
	&-\underline{2l_{A}^{y}\nabla_{+}^{2}l_{A}^{x}}]-[m_{A}^{x}(\nabla_{+}^{2}+4\frac{\partial}{\partial z})m_{B}^{y}-\underline{l_{A}^{x}(\nabla_{+}^{2}+4\frac{\partial}{\partial z})l_{B}^{y}}+2m_{A}^{x}\nabla_{+}^{2}m_{A}^{y}-\underline{2l_{A}^{x}\nabla_{+}^{2}l_{A}^{y}}]\Big\},
\end{align}
\begin{align}\label{EOM_l_AB_1}
	\frac{dl_{A}^{x}}{dt}\approx&\frac{3}{2}J_{\parallel}(\underline{m_{A}^{z}l_{B}^{y}}+l_{A}^{z}m_{B}^{y})+\frac{3}{2}J_{\perp}[(l_{A}^{z}m_{B}^{y}-\underline{l_{A}^{y}m_{B}^{z}})+(l_{B}^{z}m_{A}^{y}-\underline{l_{B}^{y}m_{A}^{z}})+4(l_{A}^{z}m_{A}^{y}-\underline{l_{A}^{y}m_{A}^{z}})]
	\nonumber
	\\
	&+\frac{3}{8}J_{\parallel}(m_{A}^{z}\nabla_{-}^{2}l_{B}^{y}+l_{A}^{z}\nabla_{-}^{2}m_{B}^{y})+\frac{3}{8}J_{\perp}\Big\{[l_{A}^{z}(\nabla_{+}^{2}+4\frac{\partial}{\partial z})m_{B}^{y}-m_{A}^{z}(\nabla_{+}^{2}+4\frac{\partial}{\partial z})l_{B}^{y}+2l_{A}^{z}\nabla_{+}^{2}m_{A}^{y}
	\nonumber
	\\
	&-2m_{A}^{z}\nabla_{+}^{2}l_{A}^{y}]-[\underline{l_{A}^{y}(\nabla_{+}^{2}+4\frac{\partial}{\partial z})m_{B}^{z}}-m_{A}^{y}(\nabla_{+}^{2}+4\frac{\partial}{\partial z})l_{B}^{z}+\underline{2l_{A}^{y}\nabla_{+}^{2}m_{A}^{z}}-2m_{A}^{y}\nabla_{+}^{2}l_{A}^{z}]\Big\},
\end{align}
\begin{align}\label{EOM_l_AB_2}
	\frac{dl_{A}^{y}}{dt}\approx&-\frac{3}{2}J_{\parallel}(\underline{m_{A}^{z}l_{B}^{x}}+l_{A}^{z}m_{B}^{x})+\frac{3}{2}J_{\perp}[(\underline{l_{A}^{x}m_{B}^{z}}-l_{A}^{z}m_{B}^{x})+(\underline{l_{B}^{x}m_{A}^{z}}-l_{B}^{z}m_{A}^{x})+4(\underline{l_{A}^{x}m_{A}^{z}}-l_{A}^{z}m_{A}^{x})]
	\nonumber
	\\
	&-\frac{3}{8}J_{\parallel}(m_{A}^{z}\nabla_{-}^{2}l_{B}^{x}+l_{A}^{z}\nabla_{-}^{2}m_{B}^{x})+\frac{3}{8}J_{\perp}\Big\{[\underline{l_{A}^{x}(\nabla_{+}^{2}+4\frac{\partial}{\partial z})m_{B}^{z}}-m_{A}^{x}(\nabla_{+}^{2}+4\frac{\partial}{\partial z})l_{B}^{z}+\underline{2l_{A}^{x}\nabla_{+}^{2}m_{A}^{z}}
	\nonumber
	\\
	&-2m_{A}^{x}\nabla_{+}^{2}l_{A}^{z}]-[l_{A}^{z}(\nabla_{+}^{2}+4\frac{\partial}{\partial z})m_{B}^{x}-m_{A}^{z}(\nabla_{+}^{2}+4\frac{\partial}{\partial z})l_{B}^{x}+2l_{A}^{z}\nabla_{+}^{2}m_{A}^{x}-2m_{A}^{z}\nabla_{+}^{2}l_{A}^{x}]\Big\},
\end{align}
and
\begin{align}\label{EOM_l_AB_3}
	\frac{dl_{A}^{z}}{dt}\approx&\underline{\frac{3}{2}J_{\parallel}[(m_{A}^{y}l_{B}^{x}+l_{A}^{y}m_{B}^{x})-(m_{A}^{x}l_{B}^{y}+l_{A}^{x}m_{B}^{y})]+\frac{3}{2}J_{\perp}[(l_{A}^{y}m_{B}^{x}-l_{A}^{x}m_{B}^{y})+(l_{B}^{y}m_{A}^{x}-l_{B}^{x}m_{A}^{y})+4(l_{A}^{y}m_{A}^{x}-l_{A}^{x}m_{A}^{y})]}
	\nonumber
	\\
	&+\frac{3}{8}J_{\parallel}[(m_{A}^{y}\nabla_{-}^{2}l_{B}^{x}+\underline{l_{A}^{y}\nabla_{-}^{2}m_{B}^{x})}-(m_{A}^{x}\nabla_{-}^{2}l_{B}^{y}+\underline{l_{A}^{x}\nabla_{-}^{2}m_{B}^{y}})]+\frac{3}{8}J_{\perp}\Big\{[\underline{l_{A}^{y}(\nabla_{+}^{2}+4\frac{\partial}{\partial z})m_{B}^{x}}-m_{A}^{y}(\nabla_{+}^{2}+4\frac{\partial}{\partial z})l_{B}^{x}
	\nonumber
	\\
	&+\underline{2l_{A}^{y}\nabla_{+}^{2}m_{A}^{x}}-2m_{A}^{y}\nabla_{+}^{2}l_{A}^{x}]-[\underline{l_{A}^{x}(\nabla_{+}^{2}+4\frac{\partial}{\partial z})m_{B}^{y}}-m_{A}^{x}(\nabla_{+}^{2}+4\frac{\partial}{\partial z})l_{B}^{y}+\underline{2l_{A}^{x}\nabla_{+}^{2}m_{A}^{y}}-2m_{A}^{x}\nabla_{+}^{2}l_{A}^{y}]\Big\}.
\end{align}
To get the equations of motion for $\bm{m}_{B}$ and $\bm{l}_{B}$, one just needs to apply (i) $A\leftrightarrow B$ and (ii) $\frac{\partial}{\partial z}\rightarrow-\frac{\partial}{\partial z}$ in Eqs. (\ref{EOM_m_AB_1})--(\ref{EOM_l_AB_3}).

At this stage, one could argue that only the \emph{underlined} terms in Eqs. (\ref{EOM_m_AB_1})--(\ref{EOM_l_AB_3}) have to be kept when discussing the \emph{linearized} dynamics of this system. The reason is that in the equilibrium $\bm{m}_{A}=\bm{m}_{B}=0$. Hence all terms quadratic in $m$ have to be ignored. Furthermore, the equilibrium positions of vectors $\bm{l}_{A}$ and $\bm{l}_{B}$ are limited to the $xy$ plane, i.e., $l_{A}^{z}=l_{B}^{z}=0$. Therefore, all terms containing a product of $l^z$ and any component of $\bm{m}$ have to be ignored. Finally, terms containing derivatives may coexist only with $l^{x,y}$, but not with $l^z$ or components of $\bm{m}$. All this limits the linearized dynamics to the underlined terms only.

Next, one could notice that the equations of motion for vectors in the sublattices $A$ and $B$ differ only by the terms containing $\pm\frac{\partial}{\partial z}$. To derive the equations, which describe the two low-energy branches of magnons, we ignore the difference in the dynamics of the $A$ and $B$ sublattices, and will proceed with the approximation when $\bm{m}_{A}=\bm{m}_{B}=\bm{m}$ and $\bm{l}_{A}=\bm{l}_{B}=\bm{l}$. In result, Eqs. (\ref{EOM_m_1})--(\ref{EOM_l_3}) are obtained.

To derive the equations of motion for the two opticlike branches, we perturb the spins on $A$ and $B$ lattices oppositely with respect to each other. With this in mind, we adopt the expansions $\bm{m}_{A/B}=\pm(\delta m^x\bm{e}_x+\delta m^y\bm{e}_y+\delta m^z\bm{e}_z)$ and $\bm{l}_{A/B}=2\tilde{S}\bm{e}_x\pm(\delta l^x\bm{e}_x+\delta l^y\bm{e}_y+\delta l^z\bm{e}_z)$, where $2\tilde{S}\bm{e}_x$ is the equilibrium N\'eel vector. Next, we substitute the expansions in $\delta m^{x,y,x}$ and $\delta l^{x,y,x}$ into Eqs. (\ref{EOM_m_AB_1})--(\ref{EOM_l_AB_3}), and keep there only the linear terms. We again neglected $\pm\frac{\partial}{\partial z}$-terms in Eqs. (\ref{EOM_m_AB_1})--(\ref{EOM_l_AB_3}) and, eventually, arrive to Eqs. (\ref{EOM_m_x})--(\ref{EOM_l_z}).

\section{Equations for $E$, $a$, and $\chi_{a,b_1,b_2}$}\label{eqs_for_ansatz}
Here the procedure is rather straightforward. We substitute the ansatz Eq. (\ref{ansatz_state}) into the eigenvalue equation $H^{+}\tilde{\psi}_1=E\tilde{\sigma}_{3}\tilde{\psi}_1$, expand $\chi_{a}$, $\chi_{b_1}$, and $\chi_{b_2}$ around $0$, and take the real parts of the equation. As a result, we get
\begin{align}\label{real_part}
	&a\Big\{[\Im(B_{k})+\Re(B_{k})(\chi_{b_1}-\chi_{a})]-C_{k}\chi_{a}+[\Im(G_{k}^{+})+\Re(G_{k}^{+})(\chi_{b_2}-\chi_{a})]\Big\}
	\nonumber
	\\
	&+\frac{1}{8a}\Big\{[\Im(B_{k})+\Re(B_{k})(\chi_{b_1}-\chi_{a})]+C_{k}\chi_{a}-[\Im(G_{k}^{+})+\Re(G_{k}^{+})(\chi_{b_2}-\chi_{a})]\Big\}\approx0,
	\nonumber
	\\
	&a\Big\{[\Im(B_{k})+\Re(B_{k})\chi_{b_2}]+C_{k}\chi_{a}+[\Im(G_{k}^{+})+\Re(G_{k}^{+})\chi_{b_1}]\Big\}
	\nonumber
	\\
	&+\frac{1}{8a}\Big\{-[\Im(B_{k})+\Re(B_{k})\chi_{b_2}]+C_{k}\chi_{a}+[\Im(G_{k}^{+})+\Re(G_{k}^{+})\chi_{b_1}]\Big\}\approx0,
	\nonumber
	\\
	&a\Big\{-[\Im(B_{k})+\Re(B_{k})(\chi_{b_1}-\chi_{a})]+C_{k}(\chi_{b_2}-\chi_{b_1})-[\Im(G_{k}^{+})+\Re(G_{k}^{+})\chi_{b_1}]\Big\}
	\nonumber
	\\
	&+\frac{1}{8a}\Big\{-[\Im(B_{k})+\Re(B_{k})(\chi_{b_1}-\chi_{a})]-C_{k}(\chi_{b_2}-\chi_{b_1})+[\Im(G_{k}^{+})+\Re(G_{k}^{+})\chi_{b_1}]\Big\}\approx0,
	\nonumber
	\\
	&a\Big\{-[\Im(B_{k})+\Re(B_{k})\chi_{b_2}]-C_{k}(\chi_{b_2}-\chi_{b_1})-[\Im(G_{k}^{+})+\Re(G_{k}^{+})(\chi_{b_2}-\chi_{a})]\Big\}
	\nonumber
	\\
	&+\frac{1}{8a}\Big\{[\Im(B_{k})+\Re(B_{k})\chi_{b_2}]-C_{k}(\chi_{b_2}-\chi_{b_1})-[\Im(G_{k}^{+})+\Re(G_{k}^{+})(\chi_{b_2}-\chi_{a})]\Big\}\approx0.
\end{align}
As for the imaginary parts, we find
\begin{align}\label{imaginary_part}
	&a[A_{k}+\Re(B_{k})+C_{k}+\Re(G_{k}^{+})]+\frac{1}{8a}[A_{k}+\Re(B_{k})-C_{k}-\Re(G_{k}^{+})]\approx E(a+\frac{1}{8a}),
	\nonumber
	\\
	&a[A_{k}+\Re(B_{k})+C_{k}+\Re(G_{k}^{+})]-\frac{1}{8a}[A_{k}+\Re(B_{k})-C_{k}-\Re(G_{k}^{+})]\approx E(-a+\frac{1}{8a}).
\end{align} 
Finally, by solving Eqs. (\ref{real_part}) and (\ref{imaginary_part}), we obtain the solution Eqs. (\ref{solution_to_eigenstates_1}) and (\ref{solution_to_eigenstates_2}).

\section{Derivation of equations of motion for $\bm{m}$ and $\bm{l}$ in the presence of deformation}\label{derivation_with_deformation}
In this appendix, we derive Eqs. (\ref{EOM_m_theta_delta_J}) and (\ref{EOM_m_phi_delta_J}). Following the discussion in Sec. \ref{macro_and_deformed} and considering the change in the exchange coupling constants according to Eq. (\ref{delta_J}), the deformed term $\sum_{\delta_1}(\delta J_{\parallel}^{\delta_1})S_{i}^{z}S_{i+\delta_1}^{y}$ in Eq. (\ref{EOM_S}) becomes
\begin{align}\label{gradient_expansion_delta_J}
	\sum_{\delta_1}(\delta J_{\parallel}^{\delta_1})S_{i}^{z}S_{i+\delta_1}^{y}=&S_{iA}^{z}S_{iB}^{y}\sum_{\delta_1}(\delta J_{\parallel}^{\delta_1})+S_{iA}^{z}[\delta J_{\parallel}^{\delta_{1,1}}\frac{\partial S_{iB}^{y}}{\partial x}+\delta J_{\parallel}^{\delta_{1,2}}(-\frac{1}{2}\frac{\partial S_{iB}^{y}}{\partial x}+\frac{\sqrt{3}}{2}\frac{\partial S_{iB}^{y}}{\partial y})
	\nonumber
	\\
	&+\delta J_{\parallel}^{\delta_{1,3}}(-\frac{1}{2}\frac{\partial S_{iB}^{y}}{\partial x}-\frac{\sqrt{3}}{2}\frac{\partial S_{iB}^{y}}{\partial y})]+\cdots
	\nonumber
	\\
	\approx&\frac{3}{2}g_1(\epsilon_{xx}+\epsilon_{yy})S_{iA}^{z}S_{iB}^{y}+\frac{3}{4}g_1S_{iA}^{z}(\bm{d}\cdot\bm{\nabla})S_{iB}^{y}.
\end{align}
Here, it was assumed that site $i$ was located on the $A$ sublattice. The vector $\bm{d}=(\epsilon_{xx}-\epsilon_{yy},-2\epsilon_{xy},0)$ describes the vector-type coupling of the deformed honeycomb lattice with the spin-wave excitations. For the case $A\rightarrow B$ the following changes should be made: (i) $A\leftrightarrow B$; (ii) $\bm{d}\rightarrow-\bm{d}$. 

The out-of-plane exchange interactions could be considered similarly to the in-plane ones. Like $g_1$, there is a new coefficient $g_2\equiv\frac{1}{c}\frac{\partial J_{\perp}}{\partial c}$, which describes the sensitivity to the \emph{inter}-plane deformation. In addition, there appears a new vector $\bm{e}$ describing the vector coupling of the out-of-plane deformations with the spin waves. In terms of the strain tensor, components of $\bm{e}$ could be found as follows: $\bm{e}=\big(-(\epsilon_{xx}-\epsilon_{yy})+4\epsilon_{xz},2\epsilon_{xy}+4\epsilon_{yz},\allowbreak2(\epsilon_{xx}+\epsilon_{yy}+2\epsilon_{zz})\big)$. Finally, we obtain a system of equations describing the spin dynamics in the presence of the lattice deformations:
\begin{align}\label{EOM_S_AB_1_delta_J}
	\frac{dS_{A/B}^{x}}{dt}\approx&(\cdots)+\frac{3}{2}g_1(\epsilon_{xx}+\epsilon_{yy})S_{A/B}^{z}S_{B/A}^{y}+\frac{3}{2}g_2(\epsilon_{xx}+\epsilon_{yy}+2\epsilon_{zz})[(S_{A/B}^{z}\bar{S}_{B/A}^{y}+2S_{A/B}^{z}\bar{S}_{A/B}^{y})-(S_{A/B}^{y}\bar{S}_{B/A}^{z}
	\nonumber
	\\
	&+2S_{A/B}^{y}\bar{S}_{A/B}^{z})]+\frac{3}{4}g_1S_{A/B}^{z}(\pm\bm{d}\cdot\bm{\nabla})S_{B/A}^{y}+\frac{3}{4}g_2[S_{A/B}^{z}(\pm\bm{e}\cdot\bm{\nabla})\bar{S}_{B/A}^{y}-S_{A/B}^{y}(\pm\bm{e}\cdot\bm{\nabla})\bar{S}_{B/A}^{z}],
\end{align}
\begin{align}\label{EOM_S_AB_2_delta_J}
	\frac{dS_{A/B}^{y}}{dt}\approx&(\cdots)-\frac{3}{2}g_1(\epsilon_{xx}+\epsilon_{yy})S_{A/B}^{z}S_{B/A}^{x}+\frac{3}{2}g_2(\epsilon_{xx}+\epsilon_{yy}+2\epsilon_{zz})[(S_{A/B}^{x}\bar{S}_{B/A}^{z}+2S_{A/B}^{x}\bar{S}_{A/B}^{z})-(S_{A/B}^{z}\bar{S}_{B/A}^{x}
	\nonumber
	\\
	&+2S_{A/B}^{z}\bar{S}_{A/B}^{x})]-\frac{3}{4}g_1S_{A/B}^{z}(\pm\bm{d}\cdot\bm{\nabla})S_{B/A}^{x}+\frac{3}{4}g_2[S_{A/B}^{x}(\pm\bm{e}\cdot\bm{\nabla})\bar{S}_{B/A}^{z}-S_{A/B}^{z}(\pm\bm{e}\cdot\bm{\nabla})\bar{S}_{B/A}^{x}],
\end{align}
and
\begin{align}\label{EOM_S_AB_3_delta_J}
	\frac{dS_{A/B}^{z}}{dt}\approx&(\cdots)+\frac{3}{2}g_1(\epsilon_{xx}+\epsilon_{yy})(S_{A/B}^{y}S_{B/A}^{x}-S_{A/B}^{x}S_{B/A}^{y})+\frac{3}{2}g_2(\epsilon_{xx}+\epsilon_{yy}+2\epsilon_{zz})[(S_{A/B}^{y}\bar{S}_{B/A}^{x}+2S_{A/B}^{y}\bar{S}_{A/B}^{x})
	\nonumber
	\\
	&-(S_{A/B}^{x}\bar{S}_{B/A}^{y}+2S_{A/B}^{x}\bar{S}_{A/B}^{y})]+\frac{3}{4}g_1[S_{A/B}^{y}(\pm\bm{d}\cdot\bm{\nabla})S_{B/A}^{x}-S_{A/B}^{x}(\pm\bm{d}\cdot\bm{\nabla})S_{B/A}^{y}]
	\nonumber
	\\
	&+\frac{3}{4}g_2[S_{A/B}^{y}(\pm\bm{e}\cdot\bm{\nabla})\bar{S}_{B/A}^{x}-S_{A/B}^{x}(\pm\bm{e}\cdot\bm{\nabla})\bar{S}_{B/A}^{y}].
\end{align}
Here, $(\cdots)$ represents all the terms on the  right hand side of Eqs. (\ref{EOM_S_AB_1}), (\ref{EOM_S_AB_2}), and (\ref{EOM_S_AB_3}) without considering the deformation in the system. Again, the equations of motion for $\bar{\bm{S}}_{A/B}$ could be obtained through the exchange $\bm{S}_{A/B}\leftrightarrow\bar{\bm{S}}_{A/B}$ in Eqs. (\ref{EOM_S_AB_1_delta_J})--(\ref{EOM_S_AB_3_delta_J}).

In terms of the macroscopic quantities $\bm{m}_{A/B}$ and $\bm{l}_{A/B}$, the equations describing the spin dynamics are
\begin{align}\label{EOM_m_AB_1_delta_J}
	\frac{dm_{A}^{x}}{dt}\approx&(\cdots)+\frac{3}{4}g_1(\epsilon_{xx}+\epsilon_{yy})(m_{A}^{z}m_{B}^{y}+\underline{l_{A}^{z}l_{B}^{y}})+\frac{3}{4}g_2(\epsilon_{xx}+\epsilon_{yy}+2\epsilon_{zz})[(m_{A}^{z}m_{B}^{y}-m_{A}^{y}m_{B}^{z})-(\underline{l_{A}^{z}l_{B}^{y}-l_{A}^{y}l_{B}^{z}})]
	\nonumber
	\\
	&+\frac{3}{8}g_1[m_{A}^{z}(\bm{d}\cdot\bm{\nabla})m_{B}^{y}+l_{A}^{z}(\bm{d}\cdot\bm{\nabla})l_{B}^{y}]+\frac{3}{8}g_2\Big\{[m_{A}^{z}(\bm{e}\cdot\bm{\nabla})m_{B}^{y}-m_{A}^{y}(\bm{e}\cdot\bm{\nabla})m_{B}^{z}]-[l_{A}^{z}(\bm{e}\cdot\bm{\nabla})l_{B}^{y}-\underline{l_{A}^{y}(\bm{e}\cdot\bm{\nabla})l_{B}^{z}}]\Big\},
\end{align}
\begin{align}\label{EOM_m_AB_2_delta_J}
	\frac{dm_{A}^{y}}{dt}\approx&(\cdots)-\frac{3}{4}g_1(\epsilon_{xx}+\epsilon_{yy})(m_{A}^{z}m_{B}^{x}+\underline{l_{A}^{z}l_{B}^{x}})+\frac{3}{4}g_2(\epsilon_{xx}+\epsilon_{yy}+2\epsilon_{zz})[(m_{A}^{x}m_{B}^{z}-m_{A}^{z}m_{B}^{x})-(\underline{l_{A}^{x}l_{B}^{z}-l_{A}^{z}l_{B}^{x}})]
	\nonumber
	\\
	&-\frac{3}{8}g_1[m_{A}^{z}(\bm{d}\cdot\bm{\nabla})m_{B}^{x}+l_{A}^{z}(\bm{d}\cdot\bm{\nabla})l_{B}^{x}]+\frac{3}{8}g_2\Big\{[m_{A}^{x}(\bm{e}\cdot\bm{\nabla})m_{B}^{z}-
	m_{A}^{z}(\bm{e}\cdot\bm{\nabla})m_{B}^{x}]
	-[\underline{l_{A}^{x}(\bm{e}\cdot\bm{\nabla})l_{B}^{z}}-l_{A}^{z}(\bm{e}\cdot\bm{\nabla})l_{B}^{x}]\Big\},
\end{align}
\begin{align}\label{EOM_m_AB_3_delta_J}
	\frac{dm_{A}^{z}}{dt}\approx&(\cdots)+\frac{3}{4}g_1(\epsilon_{xx}+\epsilon_{yy})[(m_{A}^{y}m_{B}^{x}-m_{A}^{x}m_{B}^{y})+(\underline{l_{A}^{y}l_{B}^{x}-l_{A}^{x}l_{B}^{y}})]+\frac{3}{4}g_2(\epsilon_{xx}+\epsilon_{yy}+2\epsilon_{zz})[(m_{A}^{y}m_{B}^{x}-m_{A}^{x}m_{B}^{y})
	\nonumber
	\\
	&-\underline{(l_{A}^{y}l_{B}^{x}-l_{A}^{x}l_{B}^{y})}]+\frac{3}{8}g_1\Big\{[m_{A}^{y}(\bm{d}\cdot\bm{\nabla})m_{B}^{x}-m_{A}^{x}(\bm{d}\cdot\bm{\nabla})m_{B}^{y}]+
	[\underline{l_{A}^{y}(\bm{d}\cdot\bm{\nabla})l_{B}^{x}}-\underline{l_{A}^{x}(\bm{d}\cdot\bm{\nabla})l_{B}^{y}}]\Big\}
	\nonumber
	\\
	&+\frac{3}{8}g_2\Big\{[m_{A}^{y}(\bm{e}\cdot\bm{\nabla})m_{B}^{x}-m_{A}^{x}(\bm{e}\cdot\bm{\nabla})m_{B}^{y}]-[\underline{l_{A}^{y}(\bm{e}\cdot\bm{\nabla})l_{B}^{x}-l_{A}^{x}(\bm{e}\cdot\bm{\nabla})l_{B}^{y}}]\Big\},
\end{align}	
\begin{align}\label{EOM_l_AB_1_delta_J}
	\frac{dl_{A}^{x}}{dt}\approx&(\cdots)+\frac{3}{4}g_1(\epsilon_{xx}+\epsilon_{yy})(\underline{m_{A}^{z}l_{B}^{y}}+l_{A}^{z}m_{B}^{y})+\frac{3}{4}g_2(\epsilon_{xx}+\epsilon_{yy}+2\epsilon_{zz})[(l_{A}^{z}m_{B}^{y}-\underline{l_{A}^{y}m_{B}^{z}})-(\underline{m_{A}^{z}l_{B}^{y}}-m_{A}^{y}l_{B}^{z})
	\nonumber
	\\
	&+4(l_{A}^{z}m_{A}^{y}-\underline{m_{A}^{z}l_{A}^{y}})]+\frac{3}{8}g_1[m_{A}^{z}(\bm{d}\cdot\bm{\nabla})l_{B}^{y}+
	l_{A}^{z}(\bm{d}\cdot\bm{\nabla})m_{B}^{y}]+\frac{3}{8}g_2\Big\{[l_{A}^{z}(\bm{e}\cdot\bm{\nabla})m_{B}^{y}-\underline{l_{A}^{y}(\bm{e}\cdot\bm{\nabla})m_{B}^{z}}]
	\nonumber
	\\
	&-[m_{A}^{z}(\bm{e}\cdot\bm{\nabla})l_{B}^{y}
	-m_{A}^{y}(\bm{e}\cdot\bm{\nabla})l_{B}^{z}]\Big\},
\end{align}
\begin{align}\label{EOM_l_AB_2_delta_J}
	\frac{dl_{A}^{y}}{dt}\approx&(\cdots)-\frac{3}{4}g_1(\epsilon_{xx}+\epsilon_{yy})(\underline{m_{A}^{z}l_{B}^{x}}+l_{A}^{z}m_{B}^{x})+\frac{3}{4}g_2(\epsilon_{xx}+\epsilon_{yy}+2\epsilon_{zz})[(\underline{l_{A}^{x}m_{B}^{z}}-l_{A}^{z}m_{B}^{x})-(m_{A}^{x}l_{B}^{z}-\underline{m_{A}^{z}l_{B}^{x}})
	\nonumber
	\\
	&+4(\underline{l_{A}^{x}m_{A}^{z}}-m_{A}^{x}l_{A}^{z})]-\frac{3}{8}g_1[m_{A}^{z}(\bm{d}\cdot\bm{\nabla})l_{B}^{x}+l_{A}^{z}(\bm{d}\cdot\bm{\nabla})m_{B}^{x}]+\frac{3}{8}g_2\Big\{[\underline{l_{A}^{x}(\bm{e}\cdot\bm{\nabla})m_{B}^{z}}-l_{A}^{z}(\bm{e}\cdot\bm{\nabla})m_{B}^{x}]
	\nonumber
	\\
	&-[m_{A}^{x}(\bm{e}\cdot\bm{\nabla})l_{B}^{z}-m_{A}^{z}(\bm{e}\cdot\bm{\nabla})l_{B}^{x}]\Big\},
\end{align}
and
\begin{align}\label{EOM_l_AB_3_delta_J}
	\frac{dl_{A}^{z}}{dt}\approx&(\cdots)+\underline{\frac{3}{4}g_1(\epsilon_{xx}+\epsilon_{yy})[(m_{A}^{y}l_{B}^{x}-m_{A}^{x}l_{B}^{y})+(l_{A}^{y}m_{B}^{x}-l_{A}^{x}m_{B}^{y})]+\frac{3}{4}g_2(\epsilon_{xx}+\epsilon_{yy}+2\epsilon_{zz})[(l_{A}^{y}m_{B}^{x}-l_{A}^{x}m_{B}^{y})}
	\nonumber
	\\
	&\underline{-(m_{A}^{y}l_{B}^{x}-m_{A}^{x}l_{B}^{y})+4(l_{A}^{y}m_{A}^{x}-m_{A}^{y}l_{A}^{x})]}+\frac{3}{8}g_1\Big\{[m_{A}^{y}(\bm{d}\cdot\bm{\nabla})l_{B}^{x}-m_{A}^{x}(\bm{d}\cdot\bm{\nabla})l_{B}^{y}]+[\underline{l_{A}^{y}(\bm{d}\cdot\bm{\nabla})m_{B}^{x}}
	\nonumber
	\\
	&\underline{-l_{A}^{x}(\bm{d}\cdot\bm{\nabla})m_{B}^{y}}]\Big\}+\frac{3}{8}g_2\Big\{[m_{A}^{y}(\bm{e}\cdot\bm{\nabla})m_{B}^{x}-m_{A}^{x}(\bm{e}\cdot\bm{\nabla})m_{B}^{y}]-[m_{A}^{y}(\bm{e}\cdot\bm{\nabla})l_{B}^{x}-m_{A}^{x}(\bm{e}\cdot\bm{\nabla})l_{B}^{y}]\Big\}.
\end{align}
Again, $(\cdots)$ is the short notation, which represents all the terms on the right hand side of the unperturbed Eqs. (\ref{EOM_m_AB_1})--(\ref{EOM_l_AB_3}). To obtain the equations of motion for $\bm{m}_{B}$ and $\bm{l}_{B}$, one just needs to apply (i) $A\leftrightarrow B$, (ii) $\frac{\partial}{\partial z}\rightarrow-\frac{\partial}{\partial z}$, (iii) $\bm{d}\rightarrow-\bm{d}$, and (iv) $\bm{e}\rightarrow-\bm{e}$ to Eqs. (\ref{EOM_m_AB_1_delta_J})--(\ref{EOM_l_AB_3_delta_J}).

For linearized dynamics, we may keep in the above equations the underlined terms only. Next, we drop all the terms containing $\pm\frac{\partial}{\partial z}$, $\bm{d}$, and $\bm{e}$. As we have argued previously, under this approximation, the equations for $A$ and $B$ sublattices coincide, and we will assume that $\bm{m}_{A}\allowbreak=\bm{m}_{B}=\bm{m}$ and $\bm{l}_{A}=\bm{l}_{B}=\bm{l}$. Finally, by applying the standard parametrization, we find the linearized equations for $m_{\theta}$, $m_{\phi}$, $\theta$, and $\phi$ in the presence of a lattice deformation, i.e., Eqs. (\ref{EOM_m_theta_delta_J}) and (\ref{EOM_m_phi_delta_J}) written in the main text.

\section{The $XY$ model versus the $XXZ$ model}\label{XY_model}
In this appendix, we justify the Hamiltonian (\ref{H_spins_explicit}), i.e., the $XY$-type intralayer exchange coupling of this model, by exploiting our macroscopic description and comparing it with the experimental data extracted from Ref. [\onlinecite{yuan2020dirac}]. To do it, we first consider a general spin Hamiltonian of the type $XXZ$, i.e.,
\begin{align}\label{H_spins_XXZ}
	H=&\sum_{i,\delta_1}J_{\parallel}(S_{i}^{x}S_{i+\delta_1}^{x}+S_{i}^{y}S_{i+\delta_1}^{y}+\alpha S_{i}^{z}S_{i+\delta_1}^{z})+\sum_{i,\delta_2}J_{\perp}(S_{i}^{x}\bar{S}_{i+\delta_2}^{x}+S_{i}^{y}\bar{S}_{i+\delta_2}^{y}+\beta S_{i}^{z}\bar{S}_{i+\delta_2}^{z})+\{S^{x/y/z}\leftrightarrow\bar{S}^{x/y/z}\}
\end{align}
where $\alpha$ and $\beta$ characterize the anisotropy in the intra- and interlayer couplings, respectively. Note that, $\alpha=0$ and $\beta=1$ leads to the $XY$ model we used in this paper. By following the same steps of deriving the equations of motion for the macroscopic variables as in Appendix \ref{derivation_without_deformation}, we get
\begin{align}\label{EOM_m_theta_XXZ}
	&\dot{m}_{\theta}\approx(4\tilde{S}^{2})\left(-\frac{3}{8}J_{\parallel}\nabla_{-}^{2}+\frac{9}{8}J_{\perp}\nabla_{+}^{2}\right)\phi,
	\nonumber
	\\
	&\dot{\phi}\approx\left(-\frac{3(1-\alpha)}{2}J_{\parallel}+\frac{9(1+\beta)}{2}J_{\perp}\right)m_{\theta};
\end{align}
and
\begin{align}\label{EOM_m_phi_XXZ}
	&\dot{m}_{\phi}\approx(4\tilde{S}^{2})\left(-\frac{3(1-\alpha)}{2}J_{\parallel}+\frac{9(1-\beta)}{2}J_{\perp}+\frac{3\alpha}{8}J_{\parallel}\nabla_{-}^{2}-\frac{9\beta}{8}J_{\perp}\nabla_{+}^{2}\right)\theta,
	\nonumber
	\\
	&\dot{\theta}\approx\left(-9J_{\perp}-\frac{3}{8}J_{\parallel}\nabla_{-}^{2}-\frac{9}{8}J_{\perp}\nabla_{+}^{2}\right)m_{\phi}.
\end{align}
for the two lowest spin-wave branches. As for the pairs $(\delta m^{y},\delta l^{z})$ and $(\delta m^{z},\delta l^{y})$, which describe other two opticlike branches, we find:
\begin{align}\label{EOM_o1_XXZ}
	\frac{d\delta m^{y}}{dt}\approx&(2\tilde{S})\left(-\frac{3(1+\alpha)}{2}J_{\parallel}+\frac{3(3-\beta)}{2}J_{\perp}-\frac{3\alpha}{8}J_{\parallel}\nabla_{-}^{2}-\frac{3\beta}{8}J_{\perp}\nabla_{+}^{2}\right)\delta l^{z},
	\nonumber
	\\
	\frac{d\delta l^{z}}{dt}\approx&(2\tilde{S})\left(3J_{\parallel}-6J_{\perp}+\frac{3}{8}J_{\parallel}\nabla_{-}^{2}-\frac{3}{8}J_{\perp}\nabla_{+}^{2}\right)\delta m^{y};
\end{align}
and
\begin{align}\label{EOM_o2_XXZ}
	\frac{d\delta m^{z}}{dt}\approx&(2\tilde{S})\left(3J_{\parallel}-3J_{\perp}+\frac{3}{8}J_{\parallel}\nabla_{-}^{2}+\frac{3}{8}J_{\perp}\nabla_{+}^{2}\right)\delta l^{y},
	\nonumber
	\\
	\frac{d\delta l^{y}}{dt}\approx&(2\tilde{S})\left(-\frac{3(1+\alpha)}{2}J_{\parallel}+\frac{3(3+\beta)}{2}J_{\perp}-\frac{3\alpha}{8}J_{\parallel}\nabla_{-}^{2}+\frac{3\beta}{8}J_{\perp}\nabla_{+}^{2}\right)\delta m^{z},
\end{align}
As a result, using Eqs. (\ref{EOM_m_theta_XXZ})--(\ref{EOM_o2_XXZ}), we obtain
\begin{align}\label{fitting_parameters}
	&v_{ax}\equiv\frac{\partial\omega_{a}}{\partial|k_{x}|}\Big|_{\bm{k}\rightarrow0}=\left(\frac{3\tilde{S}}{2}\right)\sqrt{\left(-(1-\alpha)J_{\parallel}+3(1+\beta)J_{\perp}\right)(-J_{\parallel}+3J_{\perp})},
	\nonumber
	\\
	&v_{az}\equiv\frac{\partial\omega_{a}}{\partial|k_{z}|}\Big|_{\bm{k}\rightarrow0}=\left(\frac{3\sqrt{6}\tilde{S}}{2}\right)\sqrt{\left(-(1-\alpha)J_{\parallel}+3(1+\beta)J_{\perp}\right)J_{\perp}},
	\nonumber
	\\
	&\omega_{sh}(\bm{k}\rightarrow0)=(3\sqrt{6}\tilde{S})\sqrt{\left(-(1-\alpha)J_{\parallel}+3(1-\beta)J_{\perp}\right)J_{\perp}},
	\nonumber
	\\
	&\omega_{o1}(\bm{k}\rightarrow0)=(3\sqrt{2}\tilde{S})\sqrt{\left(-(1+\alpha)J_{\parallel}+(3-\beta)J_{\perp}\right)(-J_{\parallel}+2J_{\perp})},
	\nonumber
	\\
	&\omega_{o2}(\bm{k}\rightarrow0)=(3\sqrt{2}\tilde{S})\sqrt{\left(-(1+\alpha)J_{\parallel}+(3+\beta)J_{\perp}\right)(-J_{\parallel}+J_{\perp})}.
\end{align}
We take the effective spin $\tilde{S}=1/2$ in Eq. (\ref{fitting_parameters}) and adjust the parameters $J_{\parallel}$, $J_{\perp}$, $\alpha$, and $\beta$ to fit the measurement in Ref. [\onlinecite{yuan2020dirac}]. From Figs. 3(a) and 3(e) in Ref. [\onlinecite{yuan2020dirac}], we estimate $(\omega_{o1}(\bm{k}\rightarrow0)+\omega_{o2}(\bm{k}\rightarrow0))/2\approx11.9$ meV, $v_{ax}\approx5.1$ meV, $\omega_{sh}(\bm{k}\rightarrow0)\approx5.8$ meV, and $v_{az}\approx3.9$ meV (here, the units of spin-wave velocity are indicated in meV, because we use for momenta dimensionless units). By fitting these data using Eq. (\ref{fitting_parameters}), an optimal set of the extracted parameters is found to be $J_{\parallel}\approx-4.27$ meV, $J_{\perp}\approx0.59$ meV, $\alpha\approx0.02$, and $\beta\approx0.97$, which is very close to the best fitting parameters suggested in Ref. [\onlinecite{yuan2020dirac}]. This confirms the legitimacy of the $XY$ Hamiltonian of the described system.
\end{widetext}

\bibliographystyle{apsrev}
\bibliography{MyBIB}

\begin{thebibliography}{19}
\bibitem[{not()}]{note1}
\bibinfo{note}{Note that, to fulfill this constraint, one can implement for
  vectors $\bm{m}_{A/B}$ and $\bm{l}_{A/B}$ the standard parametrization for
  their $x$, $y$, and $z$ components:
  $\bm{l}_{A/B}=2\tilde{S}\sqrt{1-(|\bm{m}_{A/B}|/2\tilde{S})^2}\Big(\cos(\theta_{A/B})\cos(\phi_{A/B}),\cos(\theta_{A/B})\allowbreak\times\sin(\phi_{A/B}),\sin(\theta_{A/B})\Big)$
  and
  $\bm{m}_{A/B}=\Big(-(m_{\theta})_{A/B}\allowbreak\times\sin(\theta_{A/B})\cos(\phi_{A/B})-(m_{\phi})_{A/B}\sin(\phi_{A/B}),-(m_{\theta})_{A/B}\allowbreak\times\sin(\theta_{A/B})\sin(\phi_{A/B})+(m_{\phi})_{A/B}\cos(\phi_{A/B}),(m_{\theta})_{A/B}\allowbreak\times\cos(\theta_{A/B})\Big)$.
  Here, $\theta_{A/B}=\pm\theta$, $\phi_{A/B}=\pm\phi$, $(m_{\theta})_{A/B}=\pm
  m_{\theta}$, and $(m_{\phi})_{A/B}=\pm m_{\phi}$. After linearization, we get
  $\delta l^{x}_{A/B}\approx0$, $\delta l^{y}_{A/B}\approx\pm2\tilde{S}\phi$,
  $\delta l^{z}_{A/B}\approx\pm2\tilde{S}\theta$, $\delta m^{x}_{A/B}\approx0$,
  $\delta m^{y}_{A/B}\approx\pm m_{\phi}$, and $\delta m^{z}_{A/B}\approx\pm
  m_{\theta}$, which are equivalent to the expansions used here.}
\end{thebibliography}

\end{document}